\documentclass[12pt,preprint]{aastex}

\def\ltsima{$\; \buildrel < \over \sim \;$}
\def\simlt{\lower.5ex\hbox{\ltsima}}
\def\gtsima{$\; \buildrel > \over \sim \;$}
\def\simgt{\lower.5ex\hbox{\gtsima}}

\newcommand{\grs}{{GRS~1915+105}}
\newcommand{\gxp}{GX~13+1}
\newcommand{\gxm}{GX~5--1}
\newcommand{\gxz}{GX~340+0}
\newcommand{\etal}{{et al.}}

\newcommand{\chandra}{{\it Chandra}}
\newcommand{\Chandra}{{\it Chandra}}
\newcommand{\nh}{$N_{\rm H}$}
\newcommand{\no}{$N_{\rm O}$}

\newcommand{\nfe}{$N_{\rm Fe}$}
\newcommand{\sio}{SiO$_2$}

\lefthead{Ueda et al.}
\righthead{Galactic Interstellar Medium}

\begin{document}

\title{Study of the Galactic Interstellar Medium from High Resolution
X-Ray Spectroscopy: X-Ray Absorption Fine Structure and Abundances of
O, Mg, Si, S, and Fe}

\author{Yoshihiro Ueda, Kazuhisa Mitsuda, Hiroshi Murakami}
\affil{Institute of Space and Astronautical Science, 
3-1-1 Yoshinodai, Sagamihara, Kanagawa 229-8510, Japan}
\email{ueda@astro.isas.jaxa.jp}

\and

\author{Kyoko Matsushita}
\affil{Department of Physics, Tokyo University of Science, 1-3 Kagurazaka,
                 Shinjuku, Tokyo 162-8601, Japan}

\begin{abstract}

We study the composition of the Galactic interstellar medium (ISM)
toward the Galactic center region ($5^\circ<|l|<20^\circ$) by utilizing
X-ray absorption features of three bright low-mass X-ray binaries
(LMXBs), \gxp, \gxm, and \gxz, observed with the \Chandra\ HETGS. We
detect X-ray absorption fine structure (XAFS) of the Si K-edge,
characterized by a narrow and a broad absorption feature at 1846 eV and
$\approx$1865 eV, respectively. Comparison with ground experimental data
indicates that most of the ISM Si exists in the form of silicates,
although a composition of ``pure'' forsterite is ruled out. The XAFS
spectra of the sulfur K-edge indicate that a significant fraction of S
exists in the gas phase. From each source, we derive the column
densities of Mg, S, Si, and Fe from the K-edge depth and that of O (or
H) from the absorption of the continuum. The elemental abundance ratios
are found to be consistent between the three targets: the mean values of
O/Si, Mg/Si, S/Si, and Fe/Si are determined to be 0.55$\pm$0.17,
1.14$\pm$0.13, 1.03$\pm$0.12, and 0.97$\pm$0.31 solar, respectively
(90\% error in the mean value). We discuss the origins of the
overabundances of the heavy metals relative to O in the Galactic ISM by
comparison with the abundance pattern of the intracluster medium in
clusters of galaxies. Assuming that most of the Mg and Si atoms are
depleted into silicates of either the proxine or olivine family, we
estimate that the number ratio of Mg to Fe in olivine is $\simgt$1.2 and
that 17\%--43\% of the total O atoms in the ISM must be contained in
silicate grains.

\end{abstract}

\keywords{dust, extinction --- ISM: abundances --- X-rays: binaries ---
techniques: spectroscopic --- X-rays: ISM}

\section{Introduction}

The composition of the Galactic interstellar medium (ISM) contains key
information for understanding the formation and evolution of the Galaxy. 
In particular, the elemental abundances of heavy metals are crucial
since the ISM is in the last stage of chemical evolution of the Galactic
disk, reflecting the history of metal enrichment by supernovae (SNe). In
addition, it is of great importance to understand the chemical
compositions of dust grains and the dust-to-gas ratio in the ISM. Dust
grains determine the physical properties of the ISM and dynamics of star
formation. For reviews of ISM abundances and dust grains in the Galaxy,
see e.g., \citet{sav96} and \citet{dra03a}, respectively.

Many efforts have been made to determine reference (or ``cosmic'')
abundances of the ISM (total of dust plus gas) by observations of
stellar photospheres, such as the Sun, main sequence B-stars, and F/G
stars, and \ion{H}{2} regions, etc. However, all these measurements
could have inherent problems. Usually, solar abundances, those
measured from the photosphere of the Sun or from meteorites, are
adopted as reference values \citep{and89}, although the 4.5 Gyr old
Sun may not truly represent the present ISM compositions. Since 1994
there have been discussions that the true ISM abundances are about 2/3
the solar values \citep{sav96,sno96}, based on new measurements of
averaged stellar abundances in B-stars \citep{gie92} and metal-rich
dwarf F/G stars \citep{edv93} in the solar neighborhood. In B stars,
it is suggested that element stratification due to diffusion is
common: the measured abundances in the stellar surface do not always
represent those of the ISM from which they form
\citep{hem03}. Although metal-rich F/G stars were expected to be young
enough to reflect the recent ISM composition, \citet{edv93} found that
there is a large scatter in the age-metalicity relation of F/G stars,
indicating that ``metal rich'' stars inevitably include old
populations. In fact, \citet{sof01} calculated averaged abundances of
only the young ($\tau <$2 Gyr) F and G disk stars from \citet{edv93}
and found that they become closer to the solar abundances. The
abundances derived from collisionally excited lines in \ion{H}{2}
regions are subject to large systematic errors because of temperature
fluctuations in the nebula \citep[e.g.,][]{mat95,kin95}. It should be
noted that there exists abundance gradients decreasing with
Galactocentric distance (e.g., summarized in Table~1 of
\citealt*{chi01}), making the issue more complex.

Studies using UV absorption lines of interstellar gas by the Goddard
High-Resolution Spectrograph onboard the {\it Hubble Space Telescope}
indicate that a large fraction of Mg, Si, and Fe are depleted into
dust grains in the Galactic disk \citep[see][]{sav96}. In fact, a
large number of silicates are expected in the ISM in order to
reproduce the extinction curve in the infrared band \citep{dra84}. The
measurement of abundance patterns in dust grains can tightly constrain
the chemical composition of the dust. The dust abundances can be
measured only indirectly, however, by subtracting the gas-phase
abundances in the ISM from the total (gas+dust) cosmic
abundances. This means that the results strongly depend on the cosmic
abundances assumed. Dust models claim that the infrared 9 and 18
$\mu$m absorption structure of silicates, due to the transitions of
the bending and stretching modes of the S-O bond, is too strong to be
explained by a sub-solar abundance of Si
\citep{sno96,mat96}. \citet{mat98} found that the problem can be
avoided by introducing ``fluffy'' dust grains, which is not supported
by a recent X-ray observation of the dust-scattered halo around
GX~13+1 \citep*{smi02}, however. The missing Si (and C) problem will
be significantly relaxed if the true ISM abundances are closer to
solar \citep{sof01}. Thus it is an urgent task to establish standard
abundances for understanding the properties of the ISM.

High-resolution X-ray absorption spectroscopy of bright Galactic sources
provides us with a powerful technique for constraining the properties of
the ISM. Unlike UV absorption lines, one can measure the metal
abundances in both the gas and the dust phase from the edge depth,
leading to a direct determination of the ISM abundances based on simple
physics. In addition, X-ray absorption fine structure (XAFS) can be used
as a diagnostic of the chemical state of an element. Analysis of XAFS,
including X-ray absorption near edge structure (XANES) and extended
X-ray absorption fine structure, could constrain not only the
dust-to-gas ratio but also the composition of dust material. Since the
launch of \chandra\ in particular, which carries transmission grating
spectrometers, extensive studies have been made using absorption
structures of the oxygen and neon K-edge using bright X-ray binaries
with hydrogen column densities of several $10^{21}$ cm$^{-2}$ (e.g.,
X~0614--091: \citealt{pae01}; Cyg X-1: \citealt{sch02}; Cyg X-2:
\citealt{tak02}; X Persei: \citealt{cun04}; multiple targets:
\citealt*{jue04}), following the pioneering work with the {\it Einstein}
observatory by \citet{sch86}. These sources are, however, not suitable
for studies of heavier elements such as Mg, Si, S, and Fe, as they
produce only shallow K-edge structures because of their low column
densities.

In this paper, we investigate the ISM composition of Mg, Si, S, and Fe
elements toward the Galactic center region by systematically analyzing
the X-ray absorption features in the X-ray spectra of three persistent
(neutron star) low-mass X-ray binaries (LMXBs), \gxp, \gxm, and \gxz,
which have absorption column densities of several $10^{22}$ cm$^{-2}$
\citep[see e.g.,][]{chr97,asa00}. The excellent energy resolution of the
\chandra\ HETGS \citep{can00} gives us the best opportunities to study
XAFS around the Mg to S K-edge. Combined with the absorption to the
continuum, we also constrain the abundance ratios not only for these
elements but also for O (and hence H, with a reasonable assumption for
O/H).

As the three targets are located at $5^\circ<|l|<20^\circ$ within
$|b|<1^{\circ}$ and at likely distances of $\simgt 7$ kpc from the Sun
\citep[see e.g., Table~4 of ][and references therein; summarized in our
Table~1]{hom04}, we can study the averaged properties of the ISM in the
Galactic disk at Galactocentric distances of 0.7--8.5 kpc. Recalling the
good correlation between the location and column density of Galactic
LMXBs, it is very likely that the majority of the absorption column
toward these targets is attributable to the ISM, not to the
circumstellar material intrinsic to the source. Furthermore, as we show
below, we obtain similar abundance ratios consistently from the three
targets, also supporting this idea. Thus, it is justified to assume that
the contribution of circumstellar matter is negligible (this point is
also discussed in \S~4.2). The spectral model of the continuum from
persistent LMXBs is well established to be a sum of a multi color disk
(MCD) model plus a black body component \citep{mit84}. Hence, there is
little uncertainty in modeling of the interstellar absorption, which
could otherwise be coupled to the unknown continuum shape. This is an
advantage of using a persistent neutron-star LMXB as a background
source. \gxm\ and \gxz\ are Z-sources, while \gxp\ is often called a
hybrid source as it shares characteristics of both a Z and an atoll
source \citep{has89}. Table~1 summarizes the mean radio flux density,
$K$ band magnitude, and column density toward the source estimated from
our X-ray spectral fit (see below) for each target; because of the large
extinction in the visible band no optical counterparts have been
identified yet.

The paper is organized as follows. \S~2 describes the observations and
data reduction. In \S~3 we present the analysis method and results on
the XAFS and abundances. Implications of our results are discussed in
\S~4, and the conclusion is given in \S~5. The solar and ``ISM''
abundances referred to in this paper correspond to the values by
\citet{and89} and by \citet*{wil00}, respectively: the O, Mg, Si, S,
and Fe abundances relative to H are $8.51\times10^{-4}$,
$3.80\times10^{-5}$, $3.55\times10^{-5}$, $1.62\times10^{-5}$, and
$4.68\times10^{-5}$ in solar abundances, and $4.90\times10^{-4}$,
$2.51\times10^{-5}$, $1.86\times10^{-5}$, $1.23\times10^{-5}$, and
$2.69\times10^{-5}$ in ``ISM'' abundances.\footnote{These solar and
``ISM'' abundances are available with the ``abund angr'' and ``abund
wilm'' command on XSPEC (version 11.2.0), respectively.}

\section{Observations and Data Reduction}

Table~2 summarizes the observation log of the three targets. We
analyzed the data with CIAO version 3.0.1 and CALDB version 2.23
according to the standard procedures. The HETGS consists of two types
of grating, the HEG (1--8 keV) and the MEG (0.5--4 keV). Only the
first order events are utilized for the spectral analysis. To
determine the zero-th point position in the image as accurately as
possible, we calculate the mean crossing point of the zeroth order
readout trace and the tracks of dispersed HEG and MEG events. The
accuracy of the absolute wavelength scale can be verified by utilizing
the sharp instrumental absorption feature at 1839 eV from polysilicon
in the CCD gate structure. We estimate the final accuracy to be about
0.002 \AA\ (HEG), and take this into account as the systematic error
when discussing results relying on the absolute energy.

Because these targets are bright, we apply pile-up correction to the
spectra with the method of \citet{ued04} applied to the same HETGS
data of \gxp. Namely, we multiply the count in each spectral bin by
the correction factor, $1/(1-f) \simeq 1 + a R$, where $R$ is the
count rate of the pixel in units of counts per pixel per frame and $a$
= 7.5 is adopted\footnote{\chandra\ Proposer's Observatory Guide
Rev.3.0 p.206}. The correction factor in the K-edge regions of
interest to us is at maximum in the MEG spectra around 2.6 keV, and is
found to be $\approx$10\%, $\approx$20\%, and $\approx$3\% for \gxp,
\gxm, and \gxz, respectively. We add a 3\% systematic error in each
bin of the spectrum, considering possible uncertainties in the
relative effective-area calibration and in the pile-up correction. We
sum up the pile-up corrected spectra and the energy responses from the
$+1$ and $-1$ orders. The spectral analysis is performed with XSPEC
version 11.2.0, with which we use the $\chi^2$ minimization technique
to find the best-fit parameters and their errors.

\section{Analysis and Results}

\subsection{Analysis Method}

The goal of our analysis is (1) to derive the absorption column
density of abundant elements, O, Mg, Si, S, and Fe (and hence their
abundance ratios), and (2) to measure the XAFS, aiming at constraining
the chemical state of the elements. Because our targets are heavily
absorbed with hydrogen column densities of \nh\ $\simeq$
(2--5)$\times10^{22}$ cm$^{-2}$ (assuming solar abundances), the
column densities of Mg, Si, S, and Fe can be derived by measuring the
depth of the K-edge. For elements lighter than Mg, in particular O,
this can be estimated from the continuum absorption. In the low energy
range, the contribution from O is the largest to the X-ray attenuation
and hence can be best estimated. We can also constrain the H column
density by assuming appropriate abundance ratios.

We analyze the data in a narrow band around the K-edge region of each
element, 1.27--1.40 keV, 1.82--1.90 keV, 2.40--2.60 keV, and
6.90--7.30 keV (7.05--7.30 keV for \gxp\ to avoid a strong absorption
line feature) for Mg, Si, S, and Fe, respectively, by a simultaneous
fit to the HEG and MEG spectra ({\it local fit}). The models used to
reproduce the XAFS are described in detail in \S 3.2. We also analyze
the HEG spectrum in a wide energy range (1--7.5 keV) to determine the
continuum and absorption column density of O, taking into account the
abundances of heavier elements determined above ({\it global fit}). In
the {\it local fit}, we fix the continuum and absorption from the
other elements at the values determined from the {\it global fit} and
other {\it local fits}. Thus, these fitting processes are repeated
iteratively until the spectral parameters become self-consistent. For
the absorption cross-sections we refer to those given in
\citet{wil00}, implemented in the {\it TBvarabs} model on XSPEC. This
model is in some cases too simple to describe the XAFS, but is
sufficient to discuss the overall continuum absorption. Because we
cannot derive the abundances of other elements individually, we always
fix the abundance ratios (excluding the element of interest in the
case of the {\it local fit}) within each group of H-He, C-N-O,
Na-Mg-Al, S-Cl-Ar-Ca, and Cr-Fe-Co-Ni at the solar values.

Because a large grain can become optically thick to X-ray absorption,
the total opacity observed from the ISM is reduced from the case in
which the same material is assumed to be completely gaseous. This
effect is calculated in Appendix~A of \citet{wil00} by assuming a
simplified grain model, and is incorporated in the {\it TBvarabs}
model. Following \citet{wil00}, we take into account this effect in
our analysis by assuming a \citet{mat77} grain-size distribution,
$\frac{d n_{\rm gr}(a)}{da} \propto a^{-3.5}$, in the range of $0.025
\mu$m $\leq a \leq 0.25 \mu$m \citep{dra84}, and a density of 1 g
cm$^{-3}$. We set the depletion factors $1-\beta_z$ = 0, 0, 1, and 0
for Mg, Si, S, and Fe, respectively, based on our results obtained
below. For other elements we adopt the default values listed in
Table~2 of \citet{wil00}. The dependence of the opacity reduction on
the depletion factor is small and the uncertainties do not affect our
discussion. For example, the opacity reduction factor is $\approx$5\%
at the Mg K-edge with $1-\beta_z=0$ (the maximum case), which is much
smaller than the statistical error.

We neglect the effects from dust scattering by the ISM to obtain the
column densities of Mg, Si, S, and Fe. The depth just at the
absorption edge energy corresponds to that of the direct beam, which
suffers from not only absorption but also scattering by the dust in
the ISM. The scattering cross-section rises above the edge energy
\citep{mit90,dra03b}, increasing the apparent edge depth. Normalizing
the scattering cross section calculated by \citet{dra03b} by the
actual intensity of the X-ray halo around GX~13+1 \citep{smi02}, we
find that the absorption column densities could be overestimated by
$\approx 10$\% for Mg, $\approx 4$\% for Si, and less for S and
Fe. These are smaller than the statistical errors. The continuum shape
is more significantly affected by dust scattering, although the effect
is partially canceled out since we measure the sum of the direct beam
and a part of the scattered halo through a complex energy response of
the HETGS convolved with the spatial extent. We take this into account
in the spectral model when determining the continuum and absorption
column densities of O (and H). The resulting $N_{\rm O}$ (and $N_{\rm
H}$) are found to be $\approx 15$\% smaller than the values obtained
without scattering correction.

While determining the column densities of Mg, S, Si, and Fe from the
edge depth is straightforward, the column density of O derived from
the continuum absorption is inevitably coupled to that of other
abundant elements, especially H, Ne, and Fe (by L electrons), that
contribute to the low energy absorption as well. This means that we
have to assume the abundance ratios between these elements to obtain
\no\ (or \nh ). As reasonable assumptions, we here consider the ratios
of O/H and O/Ne to be within 0.5--2 solar. \citet{tak02} reports the
O/H ratio to be 0.70$\pm0.20$ solar in the ISM toward Cyg X-2, which
is at a Galactocentric distance roughly similar to that of the Sun,
based on the O K-edge depth and the H column density estimated from
radio and H$\alpha$ observations. Observations of \ion{H}{2} regions
suggest abundance gradients of [O/H] of $\approx$ --0.06 dex
kpc$^{-1}$ with respect to the Galactocentric distance \citep*{aff97}. 
The combination of the two arguments implies that the O abundance is
$\simeq$ 2 solar at the Galactic center. The O/Ne abundance ratio is
determined to be $0.9\pm0.4$ solar (toward Cyg~X-2) by \citet{tak02}
and 0.9$\pm$0.6 solar (toward 4U~1626--67) by \citet{sch01}. The O/Ne
ratio may be less dependent on the Galactocentric distance than O/H as
both Ne and O have similar abundance gradients with respect to H
\citep{sim95,aff97}.

Thus, we examine the difference by independently changing the
abundance ratios of O/H and O/Ne within 0.5--2 solar, and \nfe\ within
the statistical error determined in the {\it local fit}. (In the case
of \gxp, where only an upper limit of \nfe\ is obtained, we adopt Fe/O
of 0.5 solar as the lower limit of \nfe, roughly corresponding to the
case of the other two sources.) We then regard the maximum range of
\no\ (or \nh ) allowed under these uncertainties as the final
systematic error. The minimum (maximum) \no\ is obtained when we
assume 0.5 (2.0) solar for O/H and O/Ne and the maximum (minimum)
value for \nfe. The systematic error is much larger than the
statistical error obtained from each spectral fit. Similarly, we can
constrain the allowed range of \nh, although the error becomes even
larger as it more strongly depends on the assumed O/H ratio.

We find that for all the targets the continuum before absorption can be
well described by the standard model for LMXBs, the MCD model plus a
black body component \citep{mit84}, modified with local features in some
cases. In the actual fit, artificial inverse edges at 2.07 and/or 4.74
keV have been introduced, which are most likely attributable to
calibration errors (see e.g., \S~3.1 of \citealt{mil04}). From \gxp\
absorption lines from highly ionized ions of Mg, Si, S, Ar, Ca, Cr, Mn,
and Fe are detected together with an apparent broad emission-line
feature at 6.6 keV \citep{ued04}. In the spectra of \gxz, we detect an
iron K broad emission line feature that can be modeled by a Gaussian
with a $1\sigma$ width of 150 eV centered at 6.57 keV with an equivalent
width of $\approx$40 eV. These features are taken into account in the
fit (for the absorption lines of \gxp\ we use the same model as
described in \citealt{ued04}). On the other hand, no statistically
significant local feature is detected from the HEG spectrum of
\gxm. Figure~1 shows the HEG first order ``unfolded ''spectra (i.e.,
corrected for effective area) in the 1--7.5 keV band obtained by the
final {\it global fit}. Here, we assume the best-fit abundances of Mg,
S, Si, and Fe determined by the {\it local fits}, which are detailed in
\S~3.2. The contribution of each continuum component is plotted
separately. The spectral parameters are summarized in Table~3 with
statistical errors (90\% confidence level for a single parameter),
except for \no\ and \nh\ for which the systematic errors estimated above
are attached.

The iron K emission line detected from \gxz\ suggests the presence of
a reflection component, most probably from the accretion disk. It
produces the corresponding iron K-edge feature, and hence could have
an effect on the measurement of the edge depth. The line energy
suggests that the reflector must be moderately ionized, with the most
abundant iron ions being around \ion{Fe}{22}. Because we perform a
spectral fit below 7.3 keV, only the contribution of iron ions with
ionization lower than \ion{Fe}{10} \citep{ver95} is relevant. From the
line profile we observe, approximately modeled by a Gaussian, the
fraction of such ions among the total Fe ions responsible for the
emission line is estimated to be $\approx$20\%, or 8 eV in terms of
the equivalent width. Using the relation between the effective solid
angle of the reflector and the equivalent width of the accompanying
iron-K fluorescence lines by \citet{bas78}, we find that the edge
depth contributed by the reflection component is at most 10\% of the
total one we observe, which is negligible in comparison with the
statistical error. We note that in some X-ray binaries a deep iron
K-edge feature intrinsic to the source can appear as a result of
partial covering without showing any strong emission line at 6.4 keV
\citep*[e.g.,][]{tan03}. In such cases, however, a hardening of the
continuum must be observed below the K-edge energy, which is not
significantly detected in our data. Hence, we simply convert the depth
of the iron K-edge into the absorption column density. In our
discussion of Fe abundance, we do not refer to the result of \gxp,
which might truly have strong broad iron K line emission producing
significant iron K-edge features.

\subsection{The X-Ray Absorption Fine Structure}

\subsubsection{Silicon}

We detect significant XAFS around the silicon K-edge characterized by
a narrow (absorption) peak at 1846 eV and a broad peak at 1865 eV over
the simple absorption edge curve. Figure~2(a) shows the first order
HEG and MEG spectra of \gxp\ in the 1.82--1.90 keV band folded with
the detector response. The instrumental absorption features of the CCD
are examined in detail by \citet{pri98} based on ground
experiments. The deep edge at 1839 eV (more clearly seen in the HEG
data because of its twice better resolution than the MEG) corresponds
to the K-edge of silicon in metal (polysilicon), while a part of the
absorption edge at 1846 eV corresponds to that of silicon dioxides,
both from material used in the CCD gate. The dashed line in
Figure~2(a) is the best-fit when a simple edge form, as implemented in
the {\it TBvarabs} model, is employed to fit the data by excluding the
energy range of 1.84--1.87 keV, in which complex excess absorption
features are evident. To check if our result is subject to
instrumental effects, we analyze the HETGS data of Cyg X-2
(observation ID of 1016), an LMXB with a much smaller absorption, as
reference data. We confirm that the features observed here are much
larger than the calibration uncertainties and must be attributed to an
astrophysical origin.

We find that these features can be well explained by the XAFS of
silicates. \citet{li95} report silicon K-edge XAFS (or XANES) spectra of
many crystalline silicate minerals, including astrophysically important
ones such as fayalite (Fe$_2$SiO$_4$), olivine ((Mg,Fe)$_2$SiO$_4$),
forsterite (Mg$_2$SiO$_4$), and enstatite (MgSiO$_3$). Their absorption
spectra are characterized by several peaks assigned as A--G in their
Figure~2. It is seen that at least the presence of peak C, a strong
narrow peak at 1845--1847 eV, and that of the broad peak G that appears
at energies of $\approx$20 eV above peak C, are common features to
silicates. The central energy of peak C shows a chemical shift for
different silicates; the energy of fayalite, olivine, forsterite, and
enstatite is lower than that of $\alpha$-quartz by 0.2, 0.7, 1.3, and
0.1 eV, respectively\footnote{Here we only refer to the relative energy
difference given in \citet{li95}, considering a possible uncertainty in
the absolute wavelength scale; the energy of peak C in $\alpha$-quartz
listed in \citet{li95} is 1846.8 eV, which is slightly different from
the value in \citet{pri98} (1847.7 eV) that represents well the actual
energy response of the HETGS.}. According to \citet{li95} and references
therein, the peaks of A, C, E, and G are attributed to transitions
within the molecular orbit of the SiO$_4^{-4}$ cluster, while the peaks
D and F are due to the multiple scattering (MS) process from atoms more
distant than the nearby O atoms with respect to the central Si atom that
absorbs X-rays.
To compare these with our data, we have to note the fact that these
ground data are measured from crystalline minerals, not amorphous ones, 
which are expected to be the dominant form of interstellar silicates
\citep[see e.g.,][]{dra03a}. Unfortunately, we could not find
experimental XAFS data in the literature for {\it amorphous} silicates
other than \sio. The major difference is that the resonant peaks due
to the MS process will be smoothed out in amorphous silicates because
of the lack of regularity in the atomic structure. Indeed, the
comparison of the XANES spectrum between $\alpha$-quartz (crystalline
\sio) and amorphous \sio\ in \citet*{cha95} clearly shows this effect.
Nevertheless, the ground data for crystals are useful, because the
results for \sio\ imply that the energy position of peak C and the
presence of peak G do not differ between amorphous and crystalline
silicates of the same chemicals.

Considering the fact that the basic absorption features (peak C and G)
are similar among different types of silicates, we fit the observed
silicon K-edge XAFS spectra, as a first approximation, with the
absorption coefficients of amorphous \sio\ measured precisely in
ground experiments by \citet{pri98}. The best-fit model is plotted by
solid lines in Figure~2(a)-(c). As seen in this figure, our data are
well reproduced by this model, yielding a significantly better fit
than the simple edge model. We obtain almost the same column density
of Si as in the case of the edge fit excluding the 1.84--1.87 keV
range. The derived column densities are listed in Table~3 for each
target. There is weak indication in the MEG data of \gxp\ that the fit
is not perfect in reproducing the shape of peak G. Such a discrepancy,
if not instrumental, is not surprising as the precise shape depends on
the chemical composition of the silicates (see Fig.~1 of
\citealt{li95}). We do not pursue the issue in this paper, but the
absorption feature of peak G could potentially be utilized to identify
the silicate class once ground data for amorphous silicates become
available. Finally, assuming that the chemical shift of peak C in the
ISM is the same for crystalline silicates, we could constrain the
composition of the ISM silicates. From the HEG data of \gxp, the
energy shift of peak C from that of $\alpha$-quartz is constrained to
be $>$ --1.0 eV by taking account of the systematic error in the
absolute energy scale ($\approx$0.6 eV). This value does not include
that of forsterite (--1.3 eV), thus ruling out the case in which the
ISM silicates are composed of ``pure'' forsterite. By fitting the data
with two absorption models with different peak-C energies, we estimate
that forsterite cannot exceed 88\% of the total silicates in number.

\subsubsection{Sulfur}

Figure~3 shows the observed spectra in the S K-region for the three
targets. Here we take the case of \gxz\ (Figure~3b) as an example,
which has the largest column density and thus produces the deepest S
K-edge. We find that the absorption profile can be well represented by
an edge structure around 2.48 keV with a narrow absorption line below
the edge with an equivalent width of 2.4$\pm0.7$ eV. Applying the {\it
edge} model (implemented in XSPEC) plus a negative Gaussian to the HEG
data of \gxz, we determine the energies to be 2474--2490 eV (edge) and
2469.4$\pm$0.7 eV with a 1$\sigma$ width of 1.0$^{+1.5}_{-1.0}$ eV
(absorption line). Here the uncertainty in the edge energy contains
the systematic error caused by the possible presence of additional
absorption features from FeS as described below. Consistent results
are obtained from the other sources. Since the above edge energy
includes the value adopted in the {\it TBvarabs} model (2477 eV, a
theoretical value of free S atoms taken from \citealt{ver95}), we
measure the edge depth by fitting the MEG and HEG spectra of each
target with the {\it TBvarabs} model to derive the column density of
S. A negative Gaussian is also included in the fit but its parameters
are not used to constrain the column density because the
identification of the line and the velocity dispersion are essentially
uncertain. The best-fit models are plotted in Figure~3, and the
parameters are listed in Table~3.

These absorption features indicate that a significant fraction of S
exists in the gas phase. In the absorption spectrum from free atoms we
should see at least a K$\beta$ resonance line ($n=1\rightarrow3$, where
$n$ is the principal quantum number in the atomic structure) and an
edge-like structure where the transition to higher energy levels become
unresolved and are smoothly connected to the transition to the free
continuum \citep{fri91}. This picture is consistent with our
observation. Indeed, the observed absorption line energy (2469.4$\pm$0.7
eV) matches with a theoretical estimate of the K$\beta$ resonance line
energy of \ion{S}{2}, 2467--2470 eV, which is simply calculated as the
difference between the K and M2/M3 shell ionization potentials
\citep{ind98}. Because of the limited statistics and the lack of
experimental data of ``gaseous'' sulfur, however, it is not trivial to
uniquely identify the observed features. Table~4 lists experimental and
theoretical K-edge energies of the S atoms in the gas phase (not in
solids) taken from the literature. The experimental value compiled by
\citet{sev79} is originally based on the measurement of solid-state
elemental S with a correction for the work function. As noted,
theoretical values have variation of several eV between different
methods. According to the calculation by \citet{gou91} and
\citet{jun91}, the edge energy of \ion{S}{2} is higher by $\approx$10 eV
than that of \ion{S}{1}. Assuming the edge energy of \ion{S}{1} to be
2477 eV (experimental), we thus infer that the observed edge energy
(2474--2490 eV) is consistent with that of \ion{S}{2} within the errors.

The observed spectra are inconsistent with the XAFS data expected from
a pure composition of solid iron sulfides (FeS). FeS shows not only an
absorption peak at 2470 eV, but also a broader absorption peak around
2478 eV with a comparable strength \citep{sug81}, which is not evident
in our data. Furthermore, the 2470 eV absorption peak expected from
FeS is much weaker relative to the edge depth. Thus, we can rule out
the case in which nearly all of sulfur is in the form of iron sulfides
in the ISM. However, its partial contribution to the total column
density is still possible if we can attribute the 2469 eV absorption
line mostly to the gas-phase sulfur. In fact, by applying a composite
model consisting of a negative Gaussian, an edge, and an XAFS profile
of FeS modeled from the data of \citet{sug81} with independent
normalizations to our spectra, we find that the contribution of FeS
can be comparable to that of gaseous sulfur. Similar statements also
hold true for solid sulfur. The XAFS of solid sulfur shows a strong
absorption peak at 2472 eV \citep[e.g.,][]{fil93}, which is
significantly higher than 2469.4$\pm$0.7 eV, and hence cannot account
for the whole XAFS observed here. Its partial contribution cannot be
ruled out, however. We discuss the interpretation of these results in
\S~4.4.

\subsubsection{Magnesium}

For the Mg K-edge, we fit the spectra with the {\it edge} model. To
avoid systematic effects by possible excess absorption features above
the edge energy, similar to the case of Si, we exclude the range of
1.305--1.330 keV in the fit. The resulting edge depth is then
converted to the column density of Mg by using the photo-absorption
cross section by \citet{wil00}. Figure~4 show the spectra and the
best-fit models (only the MEG spectrum can be analyzed for \gxz\
because of the poor photon statistics in this energy band for the
HEG.). The edge energy is consistently determined to be 1307$\pm$2 eV
for both \gxp\ and \gxm. Accordingly, we fix the edge at 1307 eV for
\gxz. This value is lower than that of the \ion{Mg}{1} atom obtained
both experimentally (1311 eV) and theoretically (1311--1314 eV; see
Table~4). Furthermore, a close look at the \gxm\ data suggests the
presence of excess absorption features in the region of 1310--1325 eV. 
The experimental data of the magnesium K-edge XAFS of enstatite show
complex, broad absorption features at 1310--1325 eV starting from
energies somewhat lower than 1310 eV \citep{cab98}, which is
consistent with our observation. This implies that these XAFS are
indeed attributable to magnesium silicates, although we cannot rule
out the possibility that the spectrum may be (partially) contaminated
by an unresolved, narrow K$\beta$ absorption line from Mg free atoms
or ions.

\subsubsection{Iron}

The column density of Fe is determined by measuring the edge depth of
cold iron at 7.1 keV. We fit the HEG spectra in the 6.9--7.3 keV band
using the {\it TBvarabs} model. Partially because of the limited
photon statistics and energy resolution of the HETGS data in the Fe
K-band, this model gives acceptable fits in all cases (Figure~5). It
is not practical to discuss the chemical composition of Fe in detail;
we leave this issue for future studies by the {\it Astro-E2} mission.
For \gxp, we do not use the energy band below 7.05 keV to exclude the
complex, intrinsic absorption-line feature from \ion{Fe}{26}
ions. Only an upper limit on \nfe\ is obtained for \gxp; we fixed the
abundance ratio between Fe and S at the ISM value when determining the
continuum (\S~3.1).

\section{Discussion}

\subsection{Summary of the Results}

To summarize, from the XAFS data analyzed, we find the following
implications for the composition of the ISM. Si exists almost
completely in silicates, where the fraction of forsterite cannot
exceed 88\%. In contrast, a significant fraction of S exists in the
gas phase probably as \ion{S}{2} and similar ionization stages,
although a partial contribution from FeS or solid S cannot be ruled
out. Mg is likely to be in the form of magnesium silicates. All these
arguments agree with the previous studies of the ISM in the Galactic
disk; Si and Mg are mostly depleted into dust grains, while S is not
(\citealt{sav96}; for S see also \S~4.4).  These XAFS are consistently
seen in the previously published data of other sources, including
\grs\ as reported by \citet{lee02}.

Our results yield a new measurement of ISM abundances in the Galactic
disk toward the Galactocentric direction ($5^\circ<|l|<20^\circ$). We
calculate the number ratios of H/Si, O/Si, Mg/Si, S/Si, Fe/Si, and Mg/O
atoms for each source, considering both the statistical and systematic
errors. Although \nh\ is subject to large errors because of coupling
with \no, it is still useful as a first order estimate. The resulting
values relative to the solar abundance ratios are listed in
Table~5. These values can be regarded as independent of one another
except for the ratio between O and Fe, whose abundances are strongly
coupled via continuum absorption. (Strictly speaking, there are similar
couplings for the O/Mg and the O/Si ratios as well, but the effect is
negligibly small compared with the systematic error in \no). As noted,
the results for the three (or two) targets are consistent with each
other, supporting our assumption that they should reflect the mean
characteristics of the ISM between the solar system and the Galactic
center. We therefore calculate mean values from the three targets to be
used in the following discussion. In the calculation we exclude the
result of \gxz\ for Mg and \gxp\ for Fe, which are not well constrained. 
Consequently, we determine the number ratios of H/Si, O/Si, Mg/Si, S/Si,
Fe/Si, and Mg/O to be 15,000$\pm$11000 (0.55$\pm$0.39 solar),
13.3$\pm$3.9 (0.55$\pm$0.17 solar), 1.22$\pm$0.14 (1.14$\pm$0.13 solar),
0.47$\pm$0.06 (1.03$\pm$0.12 solar), 1.28$\pm$0.41 (0.97$\pm$0.31
solar), and 0.100$\pm$0.046 (2.2$\pm$1.1 solar), respectively.

\subsection{Origin of the Abundance Pattern}

We detect significant overabundance of Mg, Si, S, and (possibly) Fe
relative to O, while the abundance ratios between Mg, Si, S, and Fe
are roughly consistent with the solar abundances. To demonstrate the
overabundance visually, we plot in Figure~1 the model spectra (dotted
line) expected when the O/Si abundance ratio is assumed to be 1 solar
with the other spectral parameters unchanged. A similar trend for the
high Mg/O or Si/O ratio ($\approx$2 solar) is noticed in the
absorption spectra of Cyg X-1 \citep{sch02} and 4U~1626--67
\citep{sch01}, in which the O K-edge is detected but the column
densities of heavier elements have large statistical errors because of
their small values (\nh\ $< 10^{22}$ cm$^{-2}$). If we assume that the
solar abundances reflect the true abundances of the ISM around the
solar system, then our result suggests that heavy metals (Mg and
above) have larger abundance gradients than O toward the Galactic
center.

We note that the extreme over-abundances of Si and Fe (but not Mg and
S) reported by \citet{lee02} from an observation of \grs\ are
apparently inconsistent with our results. This is a little puzzling,
but may be explained if the results are attributed to the Si and Fe
rich circumstellar gas around \grs\ as the authors suggest in that
paper. As discussed in \citet{tak02}, however, generally the existence
of cold atoms in circumstellar gas is unlikely from the argument of
the ionization parameter, $\xi\ \equiv L/n r^2$ \citep{tar69}, where
$L$, $n$, and $r$ are the luminosity, the density of the gas, and the
distance from the X-ray source, respectively: combined with the column
density $nr \sim 10^{22}$ cm$^{-2}$, the distance becomes unreasonably
large ($r \simgt 10^{15}$ cm) compared with the binary size (typically
$\sim 10^{12}$ cm), for the gas not to be ionized by strong
X-radiation (e.g., $\xi <$10 for $L \sim 10^{38}$ erg s$^{-1}$). This
can be reconciled only when the gas has an extremely small filling
factor (e.g., when it exists as blobs). In any case, we infer that the
case of \grs\ is exceptional and do not include the result in our
discussion. Obviously, we need more examples of Galactic binaries with
similar absorptions to make more reliable arguments on the
contribution from circumstellar matter.

The observed metal abundance pattern is consistent with that of
``very'' metal rich stars (those with [Fe/H] $\simeq$ 0.3--0.5) in the
Galactic disk. According to \citet{fel98}, the [O/Fe] ratio decreases
with [Fe/H], while [Mg/Fe] and [Si/Fe] are roughly constant at the
range of [Fe/H] $>0$.  The consistency is reasonable because the
abundances of very metal rich stars are likely to reflect those in the
recent stage of the ISM, although, as mentioned in \S~1, the large
scatter in the age-metal relation \citep{edv93} would make the
interpretation nontrivial.

It is very interesting that the intracluster medium (ICM) in the
central part of clusters (or groups) of galaxies also show similar
abundance patterns, O/Si $\sim$ 0.5 solar, Fe/Si$\sim$ 1 solar, and
Mg/O $\sim$ 1.3--2.0 solar (\citealt*{mat03}; \citealt{bou03};
\citealt{tam04}). In the ICM of these systems, the abundance of Si and
Fe are more concentrated toward the cluster center than that of
O. \citet{mat03} interpret these facts through an explanation in which
the contribution of SNe~Ia becomes more important toward the cluster
center and a similar amount of Si is produced by SNe Ia along with Fe
(O can only be produced by SNe~II, which happens in the early phase of
galaxy formation.). The luminosities of SNe Ia (i.e., the total amount of
$^{56}$Ni, which eventually decays to Fe) decreases with the age of
the stellar system \citep*{iva00}. Hence, SNe~Ia in very old stellar
systems, such as cD galaxies or elliptical galaxies in the cluster (or
group) center, are expected to produce metals with a large Si/Fe
ratio.

Since the stellar system in the Galactic bulge is also an old system,
the observed abundance pattern between O, Si, and Fe may be explained
with a similar scenario i.e., significant contribution by SNe~Ia that
produce a high Si/Fe abundance ratio superposed on that by SNe~II
responsible for the production of O and Mg. In this case, however, the
observed high Mg/Si ratio (1.14$\pm$0.13 solar) may invoke another
problem because Mg, unlike Si, cannot be produced by SNe~Ia. The
observed Mg/O ratio (2.2$\pm$1.1 solar; most of the error arises from
the uncertainty in the O abundance, the best fit corresponding to the
case of O/H = 1 solar) must reflect a pure contribution from
SNe~II. It seems larger than a typical predicted value (Mg/O $\approx$
1 solar) by theoretical models \citep[e.g.,][]{nom97}, although the
elemental abundances in the ejecta of SNe~II depend on the amount of
metal of the progenitor \citep{woo95}. The contradiction is eased if
we adopt a higher O abundance (hence lower H/O, Mg/O and Si/O ratios)
within the uncertainty. Thus, to separate the contribution of SNe~Ia
and SNe~II in the Galactic ISM, a more precise determination of the O
abundance relative to heavier elements is important.

\subsection{Chemical Compositions of Silicates}

Our results of the abundance ratios between Mg, Si, and Fe, can
constrain the chemical compositions of silicates in the Galactic
ISM. In the following discussion, we make the basic assumption that
100\% of the Si and Mg atoms in the ISM are depleted into
silicates. Furthermore, we consider only two classes of silicates, the
proxine group, with the chemical formula Mg$_{x}$Fe$_{(1-x)}$SiO$_{3}$
($0 \leq x \leq 1$; $x=0$ and 1 corresponds to ferrosilite and
enstatite, respectively), and the olivine family,
Mg$_{2y}$Fe$_{2(1-y)}$SiO$_{4}$ ($0 \leq y \leq 1$; $y=0$ and 1
corresponds to fayalite and forsterite, respectively).  These are
thought to be the dominant classes of interstellar silicates \citep[see
e.g.,][]{sav96}. Introducing the parameter $\alpha$ ($0 \leq \alpha
\leq 1$), the fraction of the proxine group in the total silicates,
[i.e., ($1-\alpha$) for the olivine group], we have the relations
\begin{equation}
\alpha = 2 - {\rm (Mg/Si)_{sil}} - {\rm (Fe/Si)_{sil}}
\end{equation}
\begin{equation}
2(1-\alpha) y = {\rm (Mg/Si)_{sil}} - \alpha x,
\end{equation}
where ${\rm (Mg/Si)_{sil}}$ and ${\rm (Fe/Si)_{sil}}$ are
the number ratio of Mg and Fe to Si atoms contained in all the
silicates, respectively. From the basic assumption, ${\rm
(Mg/Si)_{sil}} = 1.22\pm0.14$, while we have only an upper limit
of ${\rm (Fe/Si)_{sil}} < 1.69$, since the true fraction of Fe
atoms in silicates is unknown. Using the value of ${\rm
(Mg/Si)_{sil}}$ in equation (1), we see that the allowed ranges of
$\alpha$ and ${\rm (Fe/S)_{sil}}$ are $0 \leq \alpha \leq 0.92$
and $0 \leq {\rm (Fe/Si)_{sil}} \leq 0.92$ with a tight
anti-correlation between the two. Equation (2) gives a strict limit
of $y \geq 0.54$ (i.e., the number ratio of forsterite to fayalite
is $\geq$1.17; the equality holds when $\alpha=0$), indicating that the
olivine family is Mg-rich. Combined with the observed number ratio of
the total O atoms to the Si atoms in the ISM (O/Si=13.3$\pm$3.9), we
can constrain the number fraction of O atoms contained in silicates to
the total O as ${\rm (O)_{\rm sil}/(O)} = \frac{4-\alpha}{\rm
O/Si} =$ 0.17--0.43 within the range of $\alpha =$ 0--1.

To proceed further, we hereafter assume a likely value of $\alpha$ by
referring to the recent result on the composition of silicates in the
ISM obtained with the {\it Infrared Space Observatory} by
\citet*{kem04}. Based on infrared spectroscopy in the 8--13 $\mu$m
wavelength range they estimate the fractions of the proxine and
olivine groups to be 15.1\% and 84.9\% by mass, respectively. This
ratio corresponds to $\alpha\simeq0.2$ by assuming $x=0.5$ and
$y=0.7$. Then, from equation (1), we obtain ${\rm (Fe/Si)_{sil}}
= 0.58\pm0.14$ and therefore, the fraction of Fe atoms contained in
silicates compared to the total Fe is constrained as ${\rm
(Fe)_{sil}/(Fe)} = 0.45\pm0.19$. Similarly, ${\rm (O)_{\rm
sil}/(O)} = 0.29^{+0.12}_{-0.07}$ for $\alpha=0.2$, where the
uncertainty comes only from that in the O/Si abundance ratio. This
value is consistent with the result that the compound fraction of O is
about 1/3 (49\% at maximum) from the analysis of O absorption edges in
the spectrum of Cyg X-2 \citep{tak02}. Hence, the maximum number
fraction of O in the form of compounds other than silicates (such as
iron oxides) is estimated to be $(49-22) = 27\%$. Finally, equation (2)
with $\alpha=0.2$ indicates the allowed range of $y$ to be 0.55--0.85,
weakly depending on $x$, which can take any value from 0 to 1. Note
that the constraint on $y$ does not conflict with the argument that
the forsterite cannot exceed 88\% of the total silicates from the
analysis of silicon K-edge XAFS (\S~3.2.1), which corresponds to the
condition $y(1-\alpha)<0.88$.

\subsection{Sulfur Composition in the ISM}

In the diffuse ISM regions, sulfur is essentially undepleted from the
gas phase; its dominant form in \ion{H}{1} regions is \ion{S}{2}
\citep{sav96}. The situation is quite different in cold molecular clouds
(with a density of $n > 10^2$--$10^4$ cm$^{-3}$), where sulfur is highly
depleted, presumably onto dust grains
\citep*[e.g.,][]{jos86,mil90,cas94}. The chemical forms of S in the
grains are uncertain, however, as well as the physical processes of how
they form in dense environments but are destroyed in the diffuse ISM
(see \citealt{ruf99} for a model). Possible candidates for the
``missing'' sulfur include solid sulfur (S$_8$) and FeS; major S-bearing
molecules such as H$_{2}$S and SO$_2$ in grain mantles cannot account
for the cosmic S abundance \citep*[e.g., ] []{pal97}. \citet{kel02}
reported the detection of FeS grains in the protoplanetary disks around
young stellar objects. This discovery is also interesting in relation to
the hypothesis proposed by \citet{bra94} that GEMS (glass with embedded
metal and sulfides) grains, silicates infested with FeS inclusions,
found in cometary interplanetary dust particles, are primitive
interstellar grains.

Even though we have determined the abundance ratio of S/Si to be
1.03$\pm$0.12 solar, the S abundance relative to H has a large
uncertainty, $2^{+3}_{-1}$ solar. With the consideration of the
abundance gradient in the Galaxy, our measurement of total (dust+gas)
sulfur abundance does not require modification of previous results on
the depletion of sulfur derived by assuming the solar abundance as the
reference. From the sulfur K-edge XAFS spectra we find that a
significant fraction of S must be contained in the gas phase, although
a partial contribution of other forms such as FeS and solid S cannot
be ruled out; to separate the contribution from different
compositions, reliable atomic data (particularly experimental data) on
the absorption spectra of sulfur ``gas'' will be helpful. We here
recall that our results reflect the averaged properties of the ISM
along the lines of sight weighted by column density from different ISM
structures at Galactocentric distances of 0.7--8.5 kpc
(\S~1). According to Table~21.2 of \citet{cox00}, the contribution to
the total column density from molecular clouds ($n \simgt 10^2$
cm$^{-3}$) is roughly comparable to that from \ion{H}{1}/\ion{H}{2}
regions at these distances (see also \citealt{dam01} for the
comparison between the \ion{H}{1} map and predicted CO map derived
from far-infrared observations). 
In the diffuse \ion{H}{1}/\ion{H}{2} regions, sulfur is mainly in
\ion{S}{2} and higher ionization stages, which can account for the
gas-phase sulfur we detected.
The remainder must be contributed from the molecular clouds. Although
it is difficult to uniquely identify the dominant chemical forms of S in
these dense environments because of the limited data quality and 
uncertainties in the atomic data, our results are consistent with the
picture that they are iron sulfides and/or solid sulfur in dust
grains.

\section{Conclusion}

We have systematically analyzed high resolution X-ray absorption
spectra of three Galactic bright sources to determine the properties
of the ISM toward the direction of the Galactic center within $(l, b)
< (20^\circ, \pm 1^\circ)$. The conclusions are summarized as follows.

1. The silicon K-edge XAFS is characterized by a narrow absorption
feature at 1846 eV and a broad one around $\approx$1865 eV. Comparison
with ground experimental data indicates that most of the ISM Si exists
as silicates. The energy of the narrow absorption line rules out a
composition of ``pure'' forsterite.

2. The sulfur K-edge XAFS consists of an edge feature at 2474--2490 eV
and a narrow absorption line at 2469.4$\pm$0.7 eV, indicating that a
significant fraction of S exists in the gas phase. A partial
contribution from iron sulfides and/or solid sulfur is possible,
however, which may be the dominant form of S trapped onto dust grains in
molecular clouds.

3. The Mg K-edge is detected at 1307$\pm$2 eV, possibly with complex,
broad absorption features in the 1310--1325 eV range. These features are
consistent with that expected from magnesium silicates, implying that Mg
atoms in the ISM are mainly contained in silicates.

4. From the K-edge depth of each element and the continuum absorption,
we have determined the O/Si, Mg/Si, S/Si, and Fe/Si abundance ratio to
be 0.55$\pm$0.17, 1.14$\pm$0.13, 1.03$\pm$0.12, and 0.97$\pm$0.31 solar,
respectively. Similar metal abundance patterns are observed in very
metal rich stars with [Fe/H] $\simeq$ 0.3--0.5 in the Galaxy and also
from the ICM in the central part of clusters (or groups) of
galaxies. Considering the fact that the Galactic bulge is an old system
similar to cD galaxies, we infer that our result may be explained by a
significant contribution from SNe~Ia that produce a high Si/Fe abundance
ratio superposed on that from SNe~II responsible for the production of O
and Mg.

5. Assuming that most of the Mg and Si atoms are depleted into
silicates of either the proxine or olivine family, we estimate that
the number ratio of Mg to Fe in olivine is $\simgt$1.2, and that
17--43\% of total O atoms in the ISM must be contained in silicate
grains.

\acknowledgments

We thank Kenji M.\ Kojima and Atsushi Ichimura for very useful
discussions on the solid state physics and atomic physics regarding
the interpretation of the XAFS. Helpful discussions with Takashi
Onaka, Issei Yamamura, Seiichi Sakamoto, and Akiko Kawamura on the
dust composition of the ISM and the properties of molecular clouds are
greatly appreciated. We also thank Chris Baluta for his careful
reading of the manuscript, Gelsomina De Stasio and Brad Frazer for
their sending us a paper about the experimental results on the XAFS of
minerals, and the anonymous referee for some useful suggestions.

\ifnum1=1
\clearpage
\begin{figure}
\epsscale{0.49}
\plotone{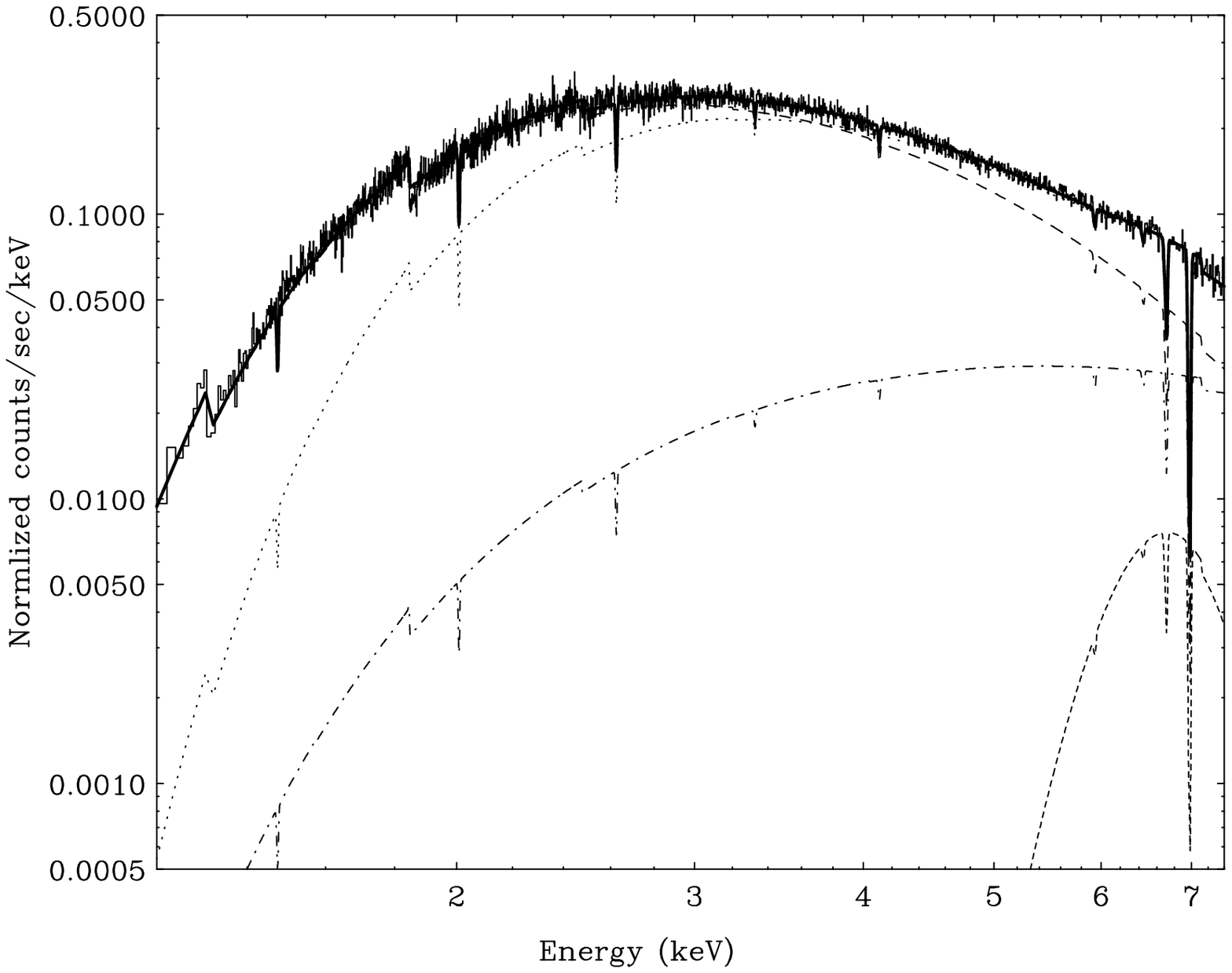}
\plotone{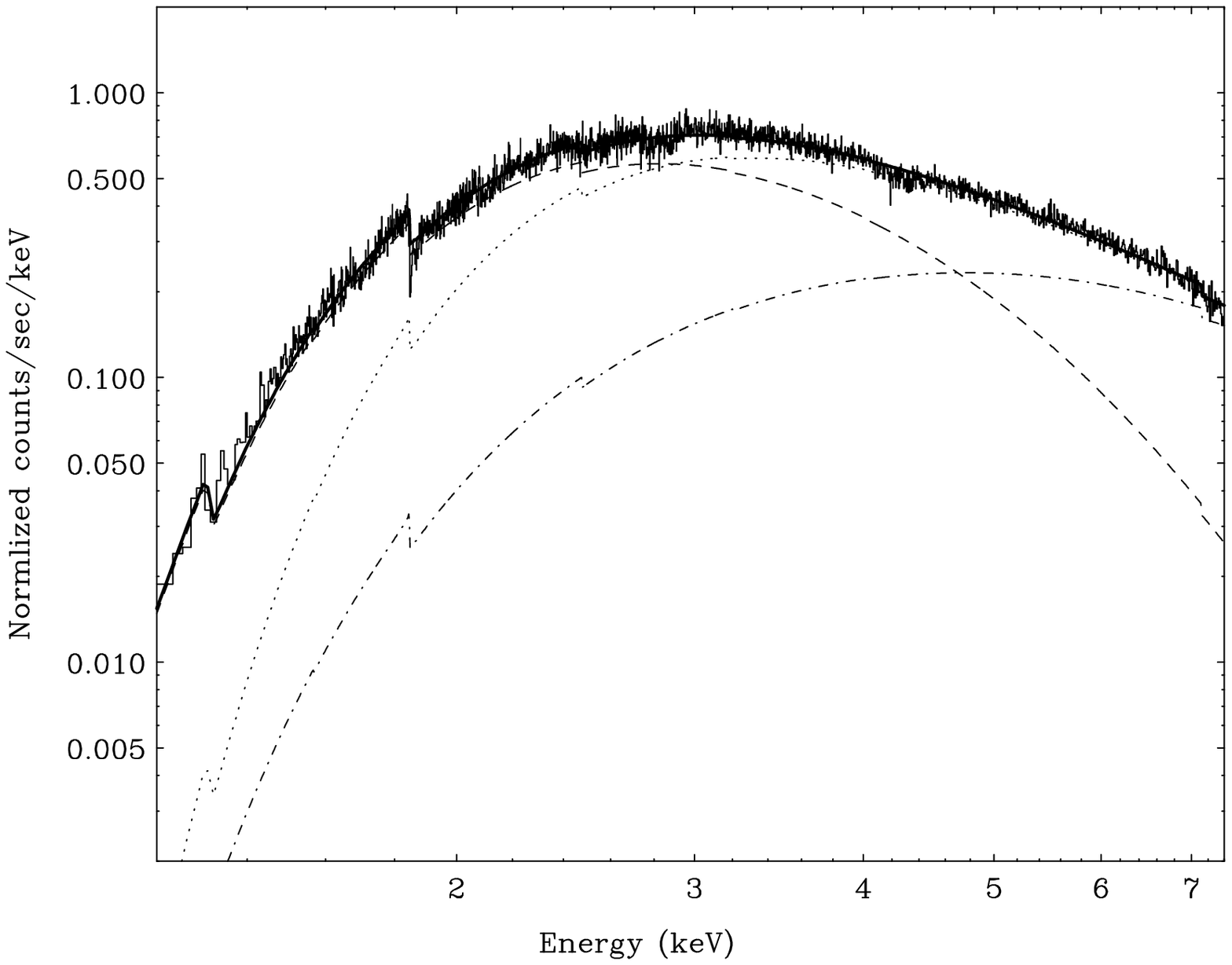}
\plotone{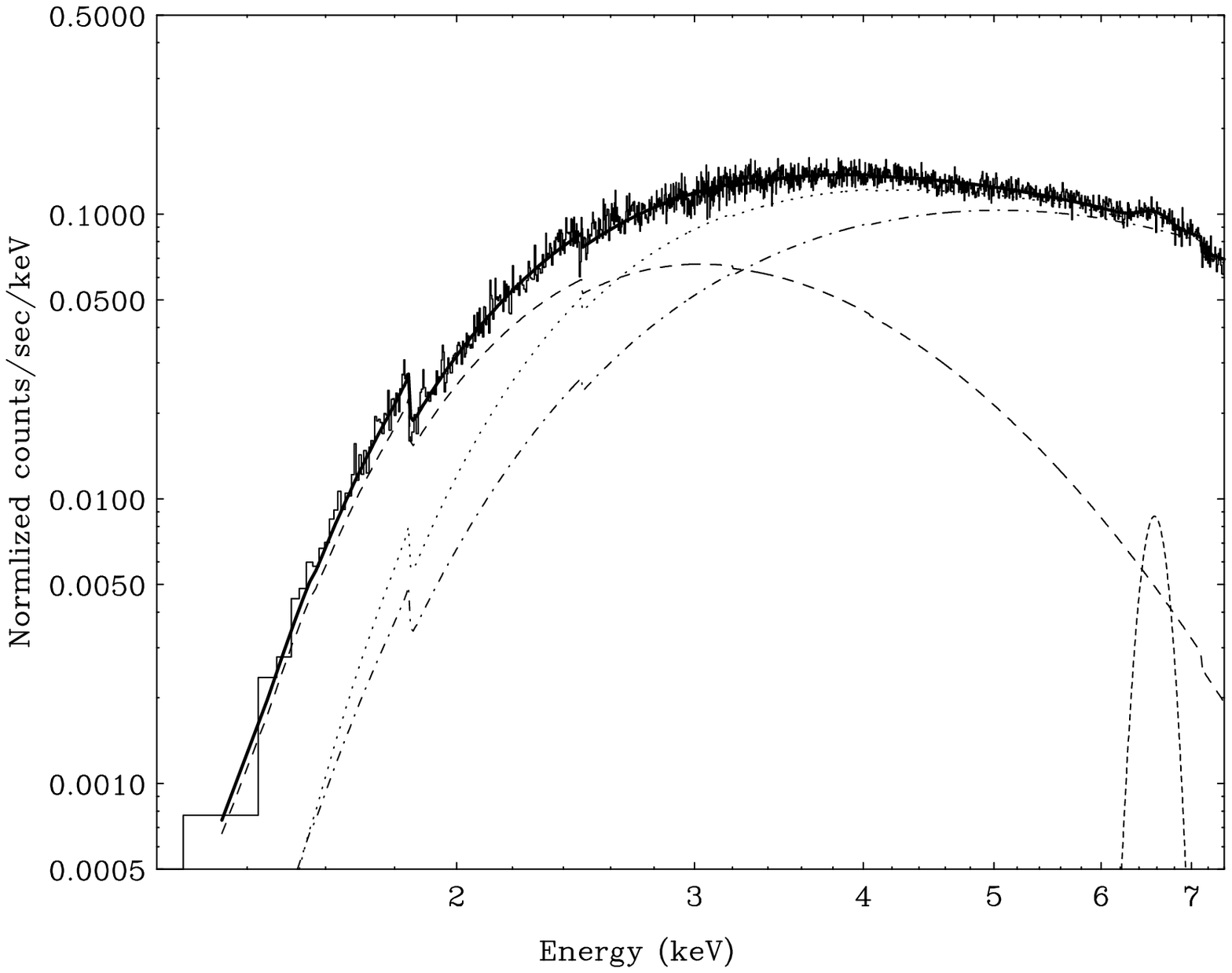}
\caption{
HEG first order ``unfolded'' spectra (i.e., corrected for the
effective area) of (a) \gxp, (b) \gxm, and (c) \gxz. The long-dashed,
dot-dashed, and short-dashed curves correspond to the MCD
component, black body, and iron K-emission line feature (not required
for \gxm), respectively.  The dotted curve shows the case in which the O/Si
abundance ratio is set to 1 solar with other parameters unchanged. The
artificial inverse edges at 2.07 and/or 4.74 keV (see text) are
excluded in these plots.
\label{fig1}}

\end{figure}

\begin{figure}
\epsscale{0.49}
\plotone{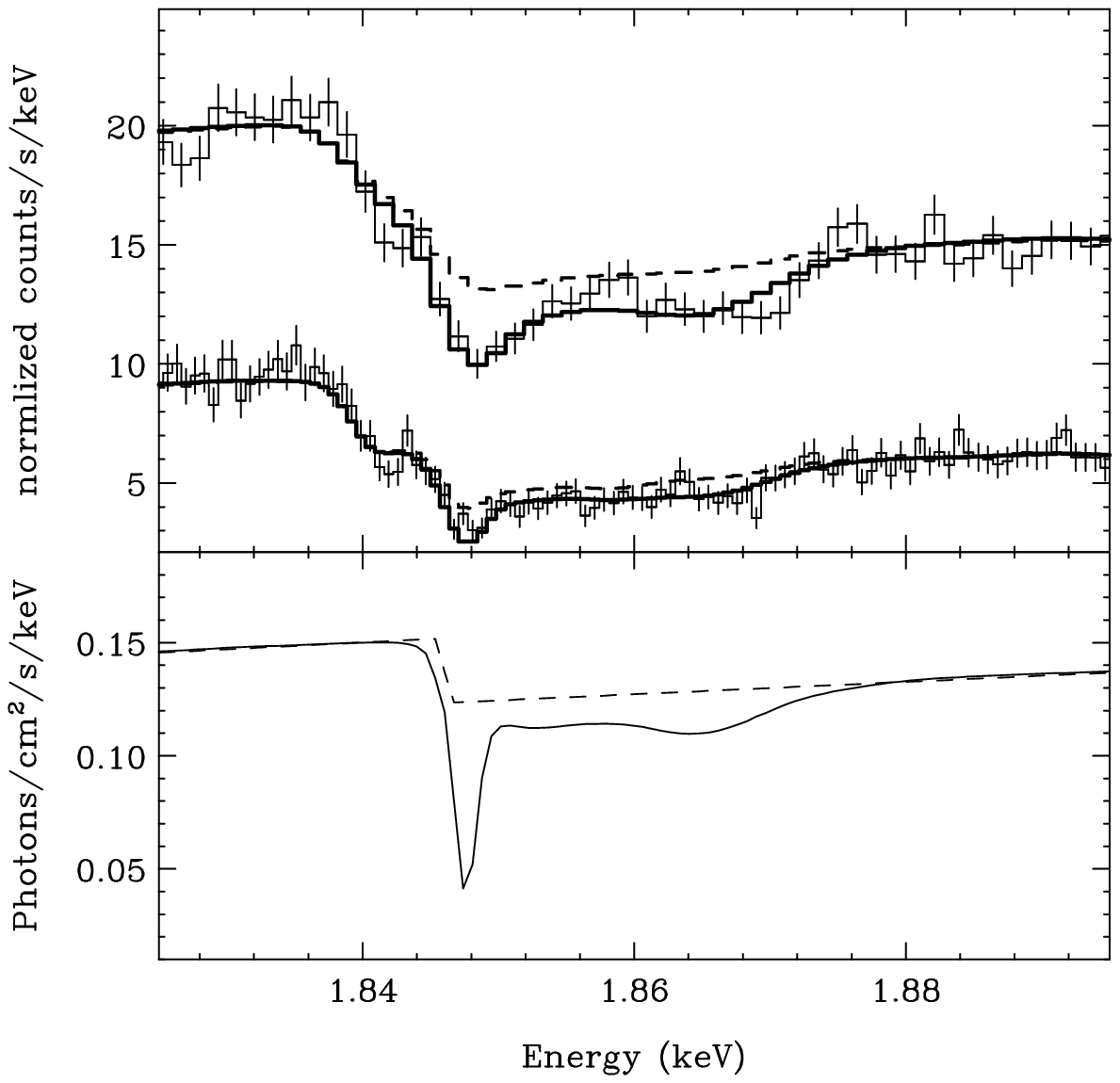}
\plotone{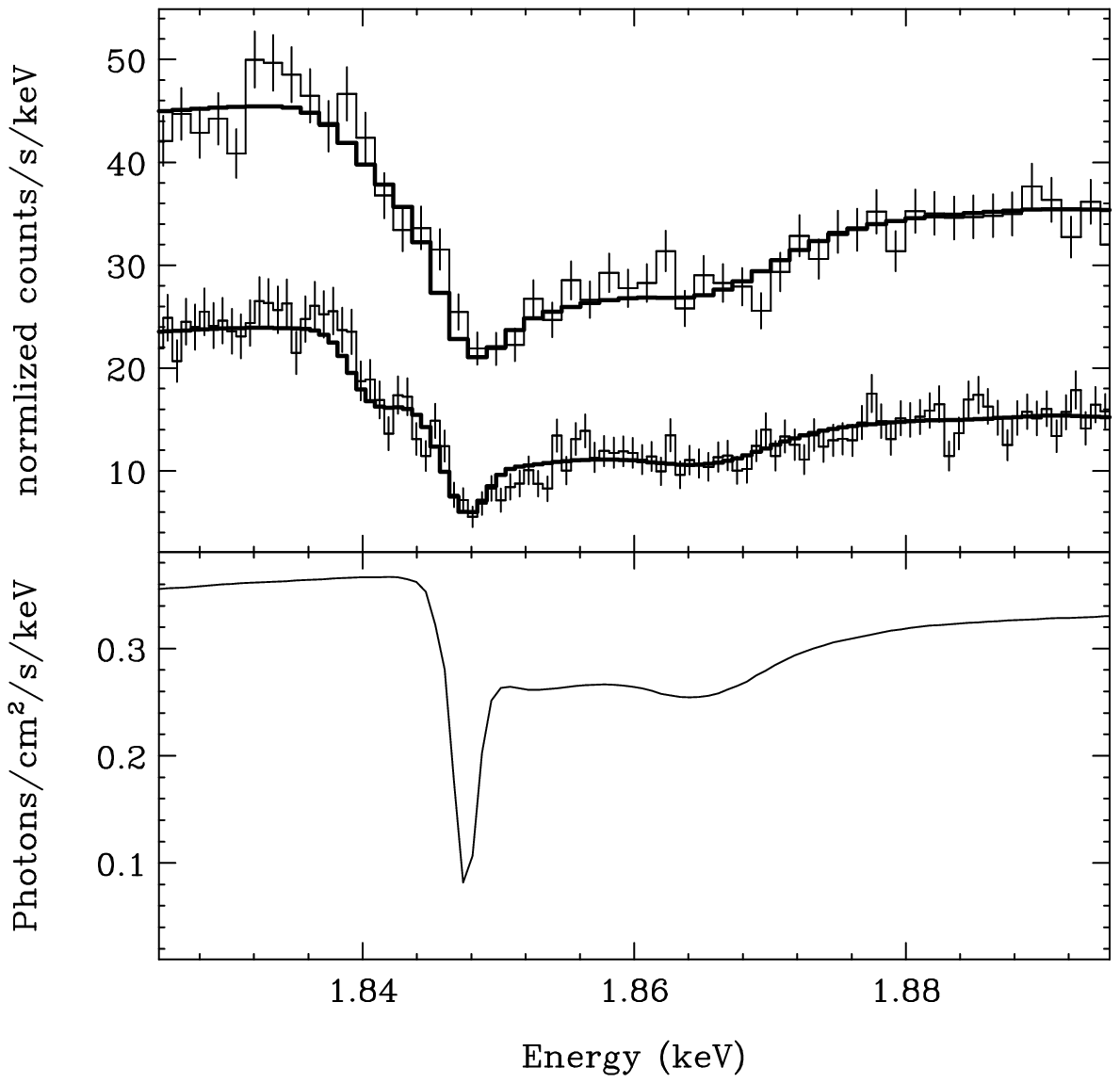}
\plotone{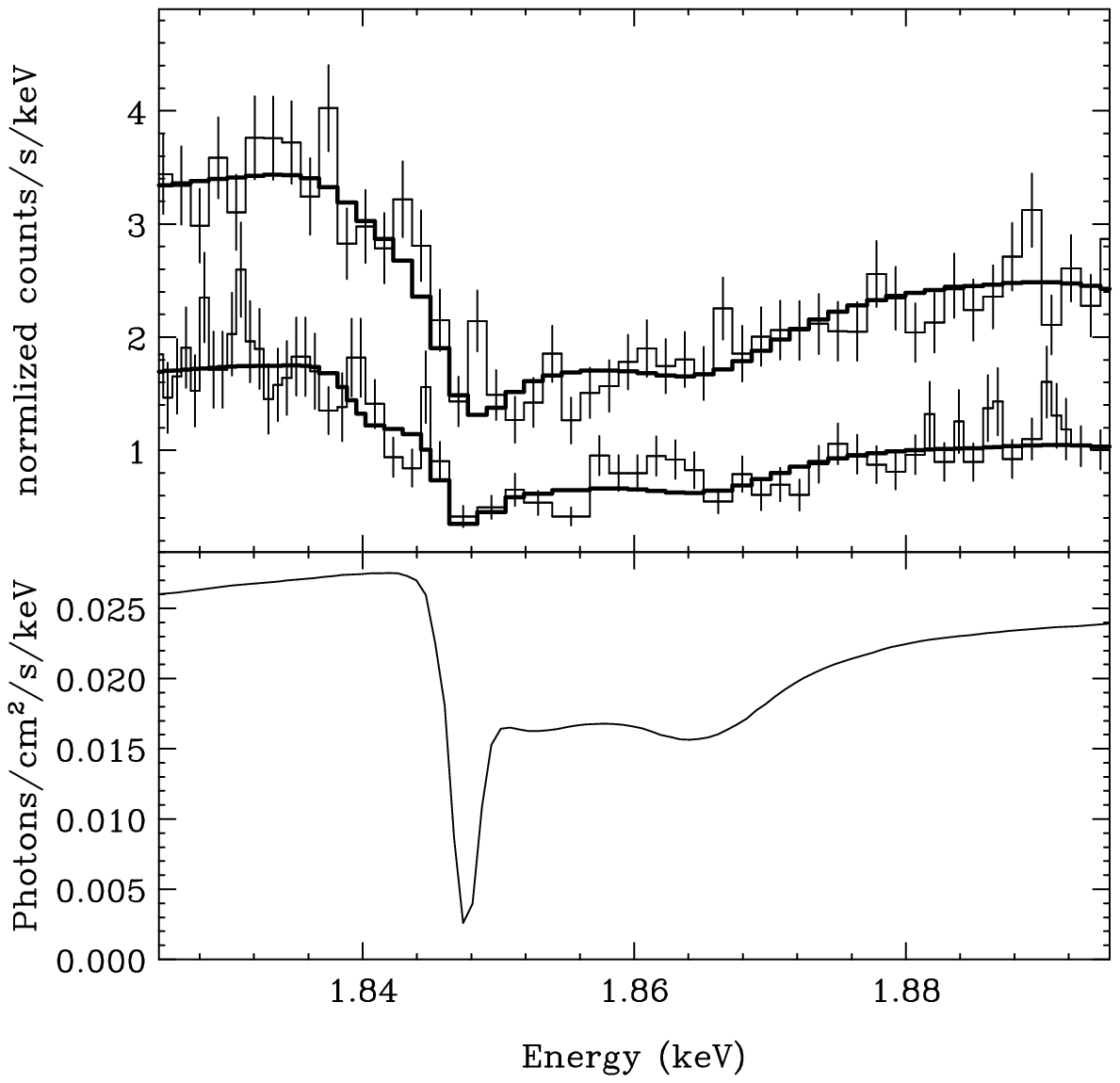}
\caption{
HEG (the lower spectrum, upper plots) and MEG first order pile-up
corrected spectra of (a) \gxp, (b) \gxm, and (c) \gxz, around the Si
K-band. The best-fit folded models (absorption by silicates) are
overplotted onto the data in the upper plots in each panel, while the
unfolded models are in the plots. The dashed lines in (a)
correspond to a simple edge model ({\it TBvarabs}) with the same
column density of Si as determined by the silicate absorption model.
\label{fig2}}
\end{figure}

\clearpage
\begin{figure}
\epsscale{0.49}
\plotone{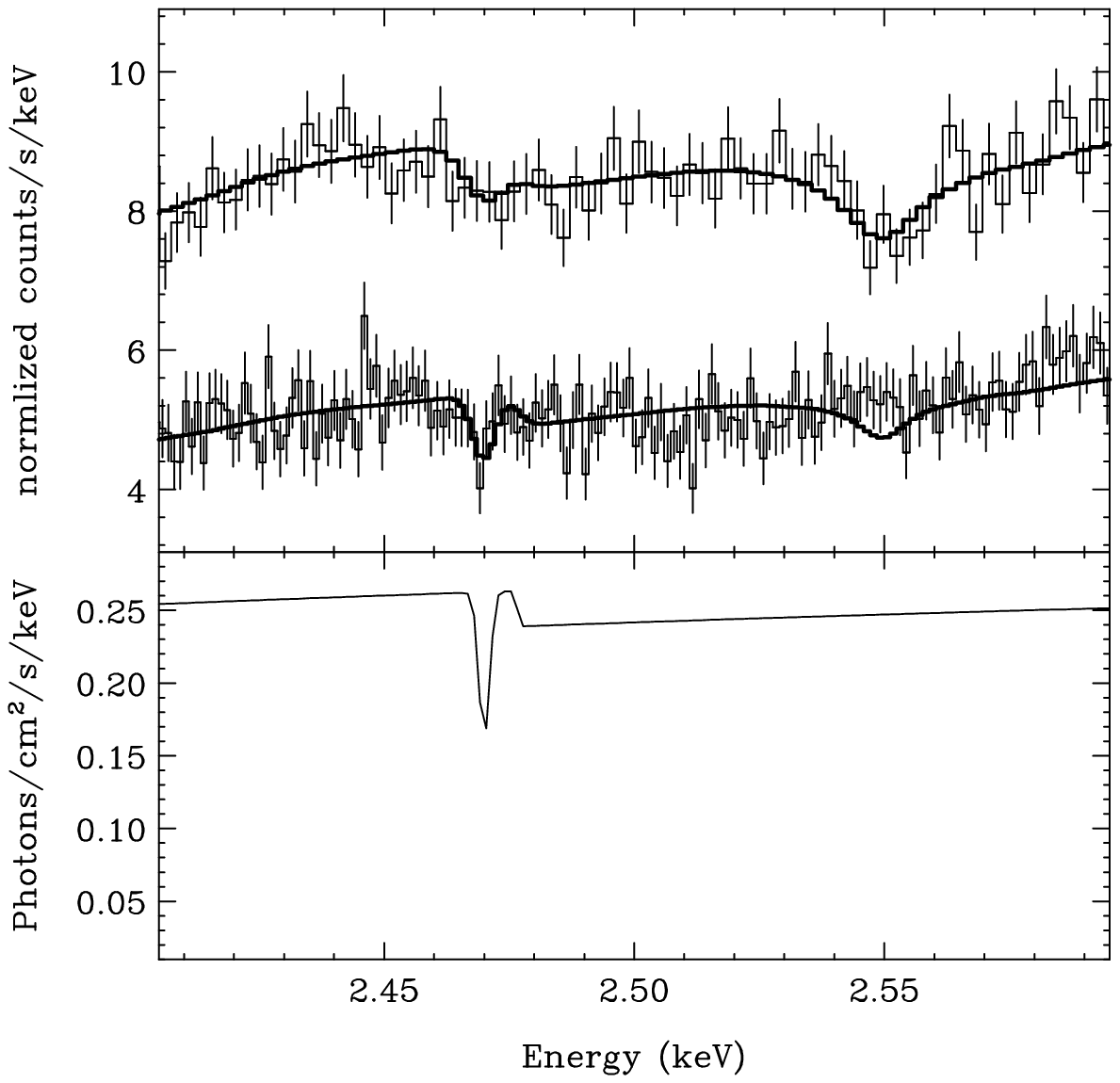}
\plotone{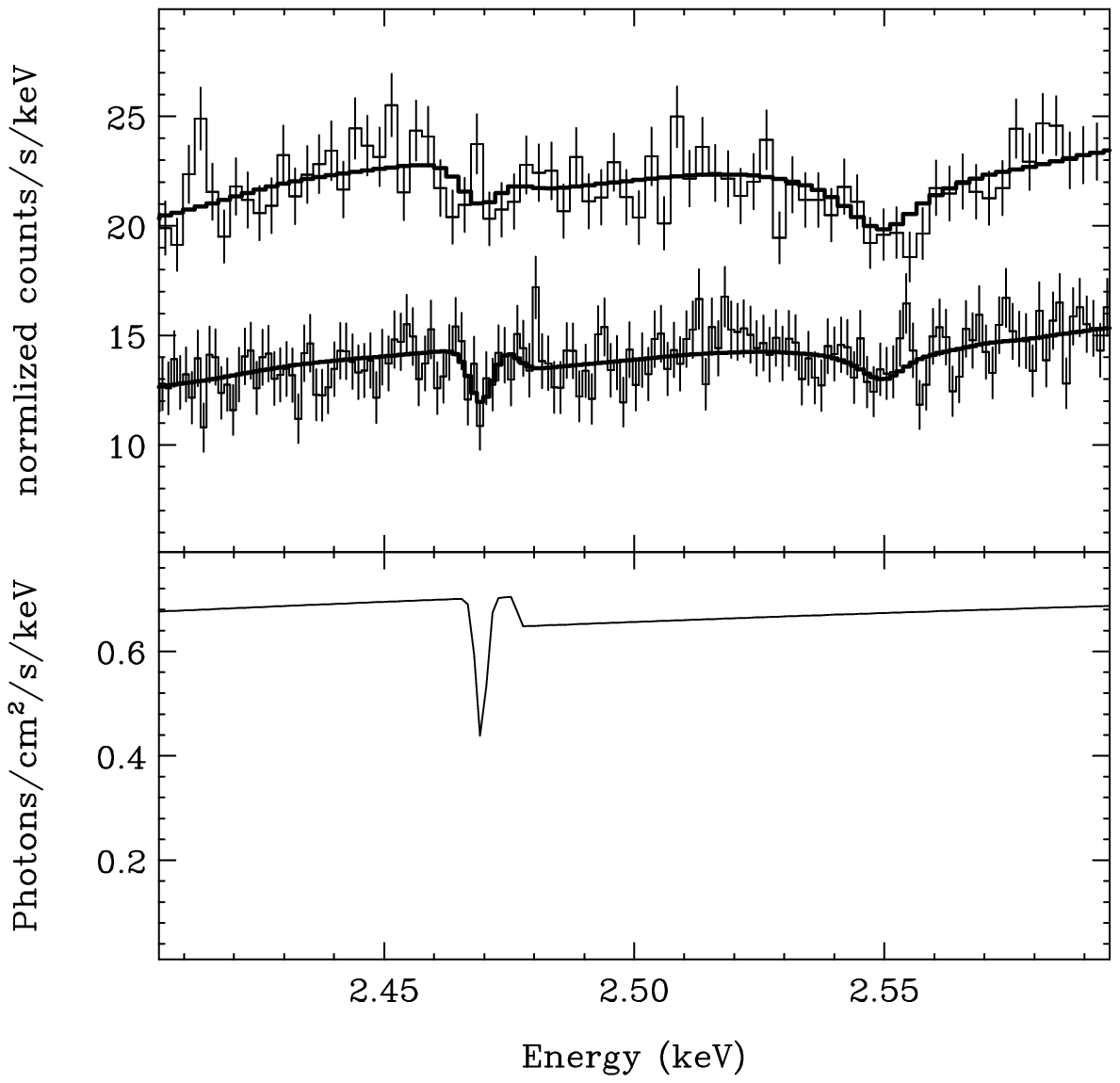}
\plotone{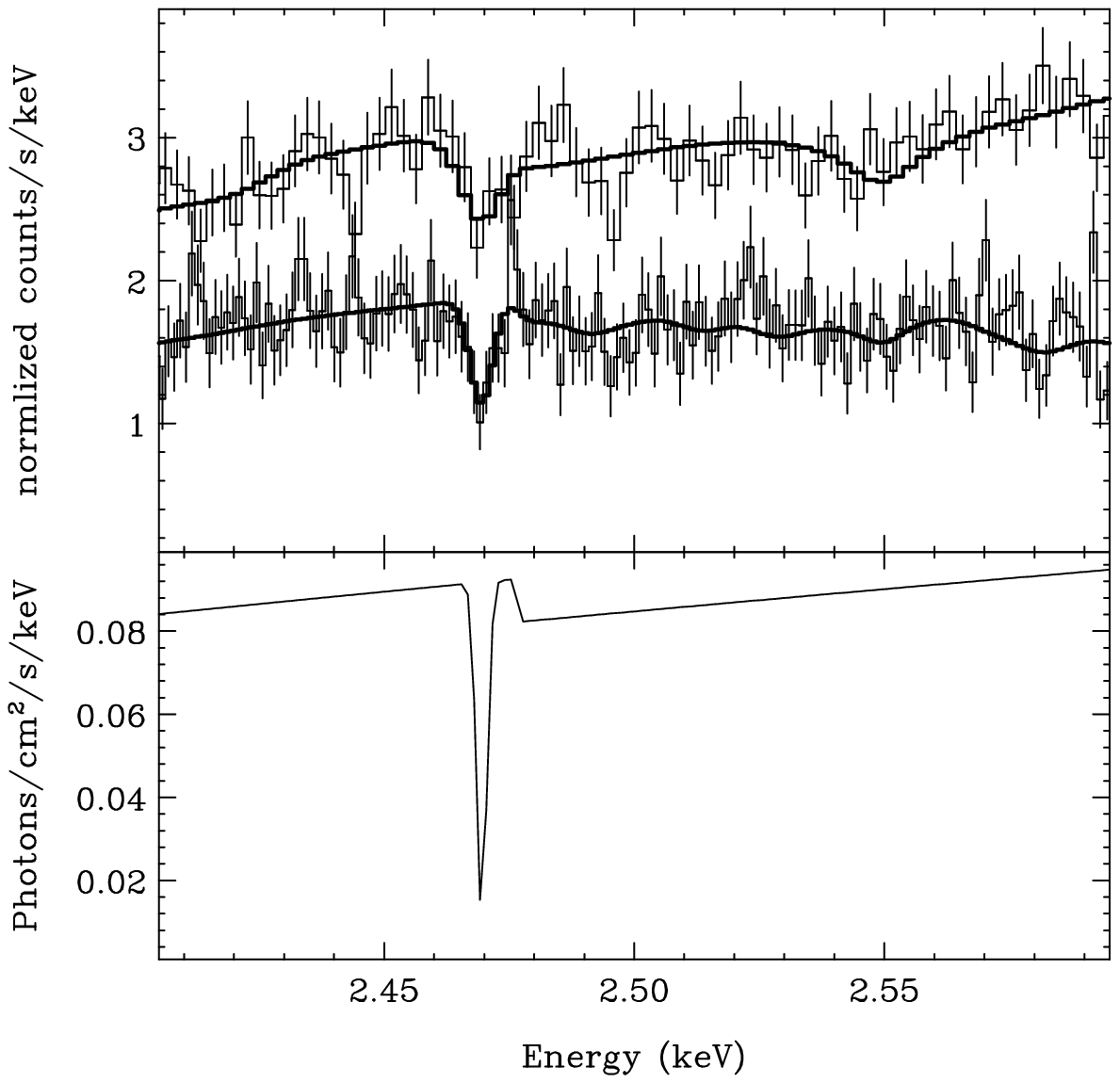}
\caption{
HEG and MEG spectra of (a) \gxp, (b) \gxm, and (c) \gxz, around
the S K-band. The best-fit folded models ({\it TBvarabs} model plus a
negative Gaussian) are overplotted onto the data in the upper plots,
while the unfolded models are in the lower plots.
\label{fig3}}
\end{figure}

\clearpage
\begin{figure}
\epsscale{0.49} 
\plotone{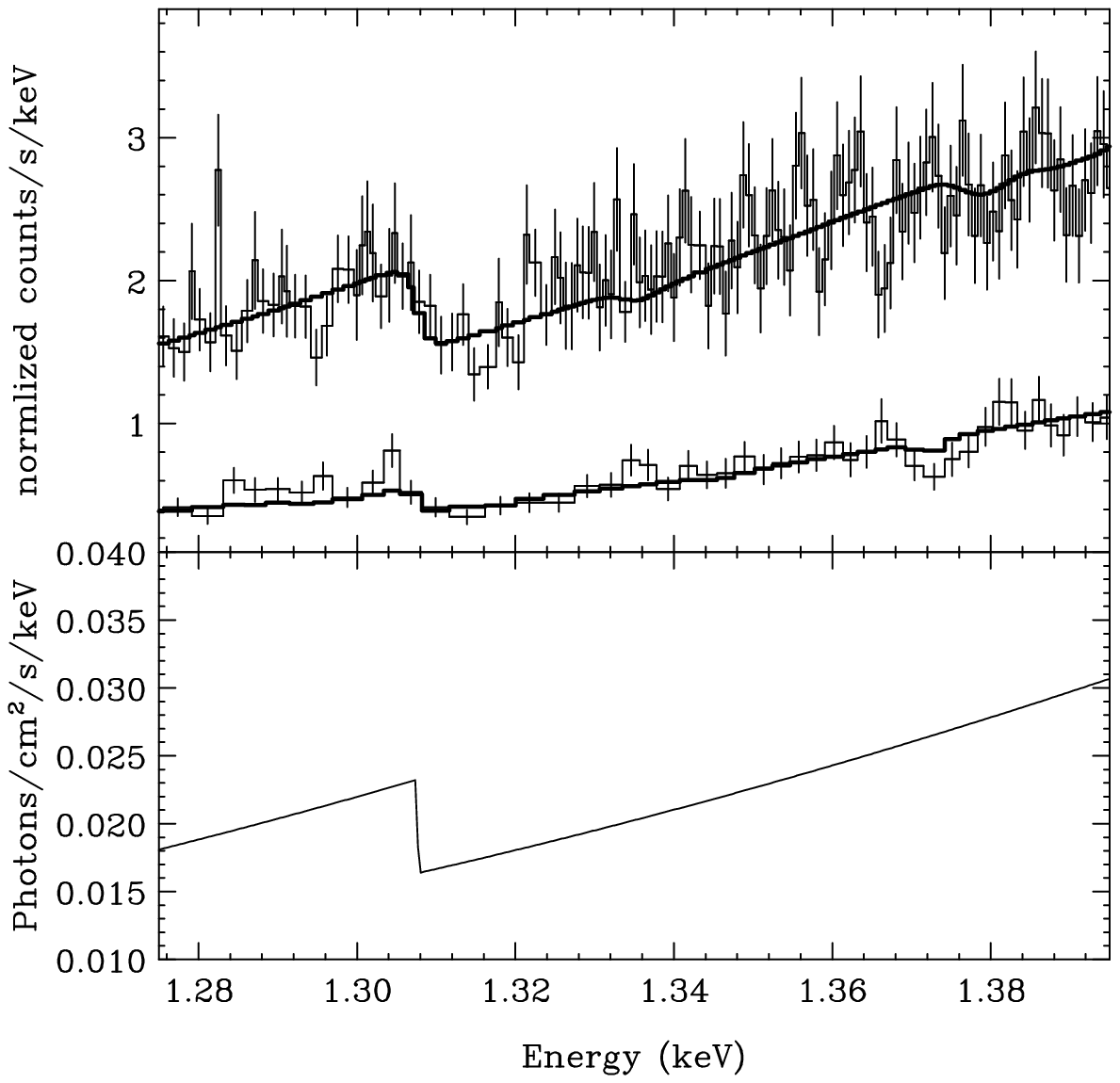}
\plotone{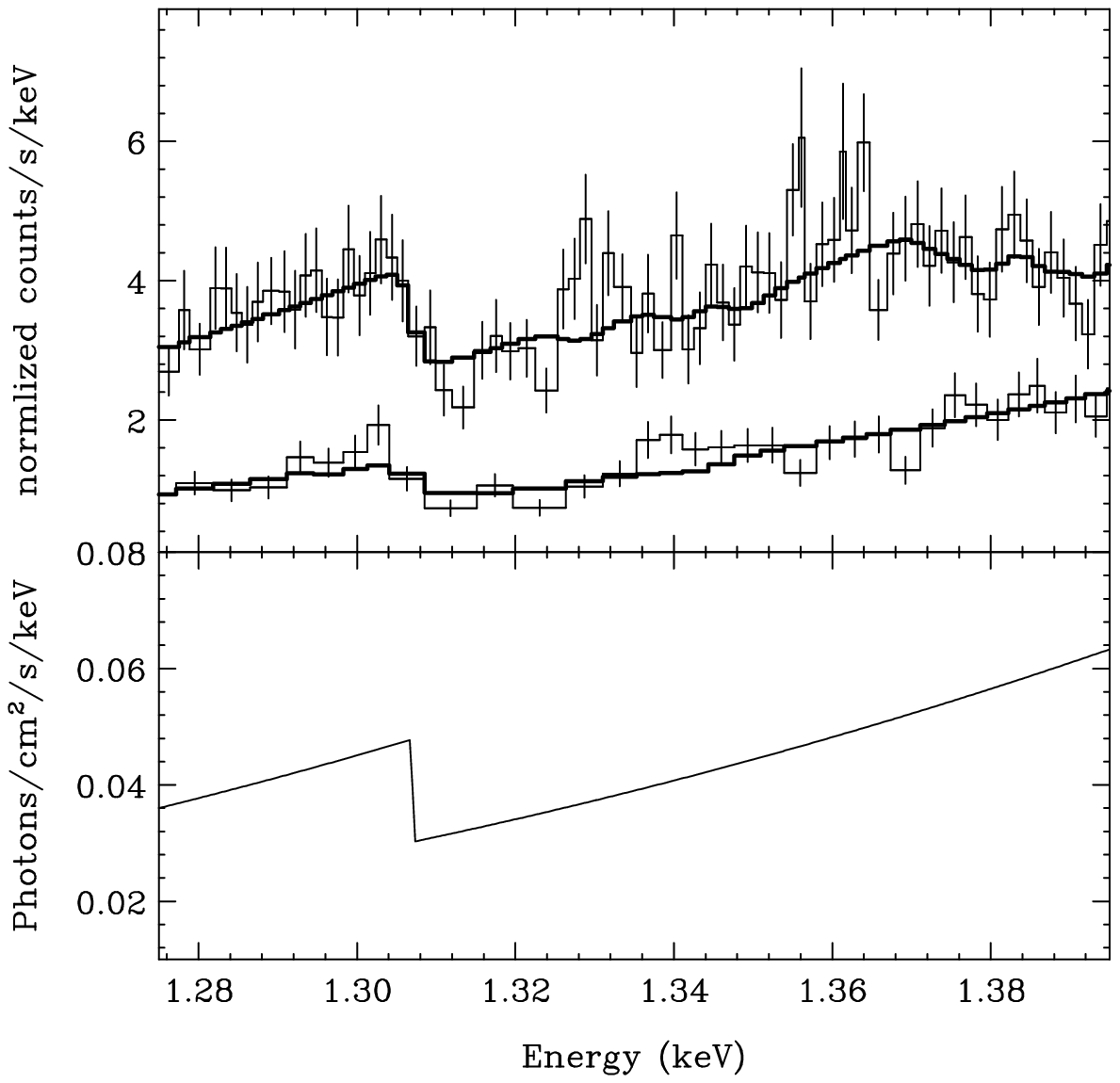}
\plotone{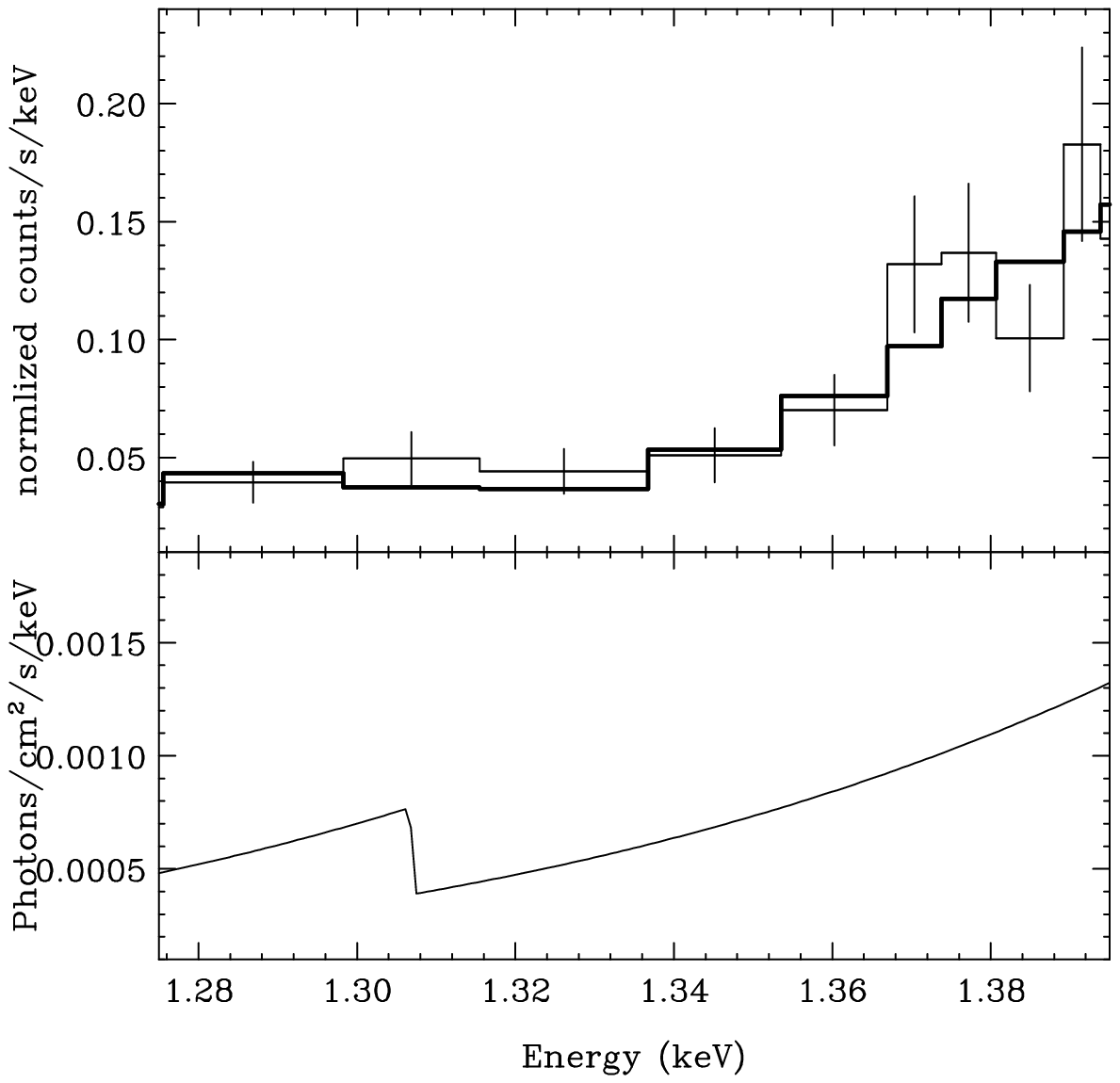}
\caption{
HEG and MEG spectra of (a) \gxp, (b) \gxm, and (c) \gxz\ (MEG
only), around the Mg K-band. The best-fit folded models ({\it edge}
model) are overplotted onto the data in the upper plots, while the
unfolded models are in the lower plots.
\label{fig4}}
\end{figure}

\clearpage
\begin{figure}
\epsscale{0.49}
\plotone{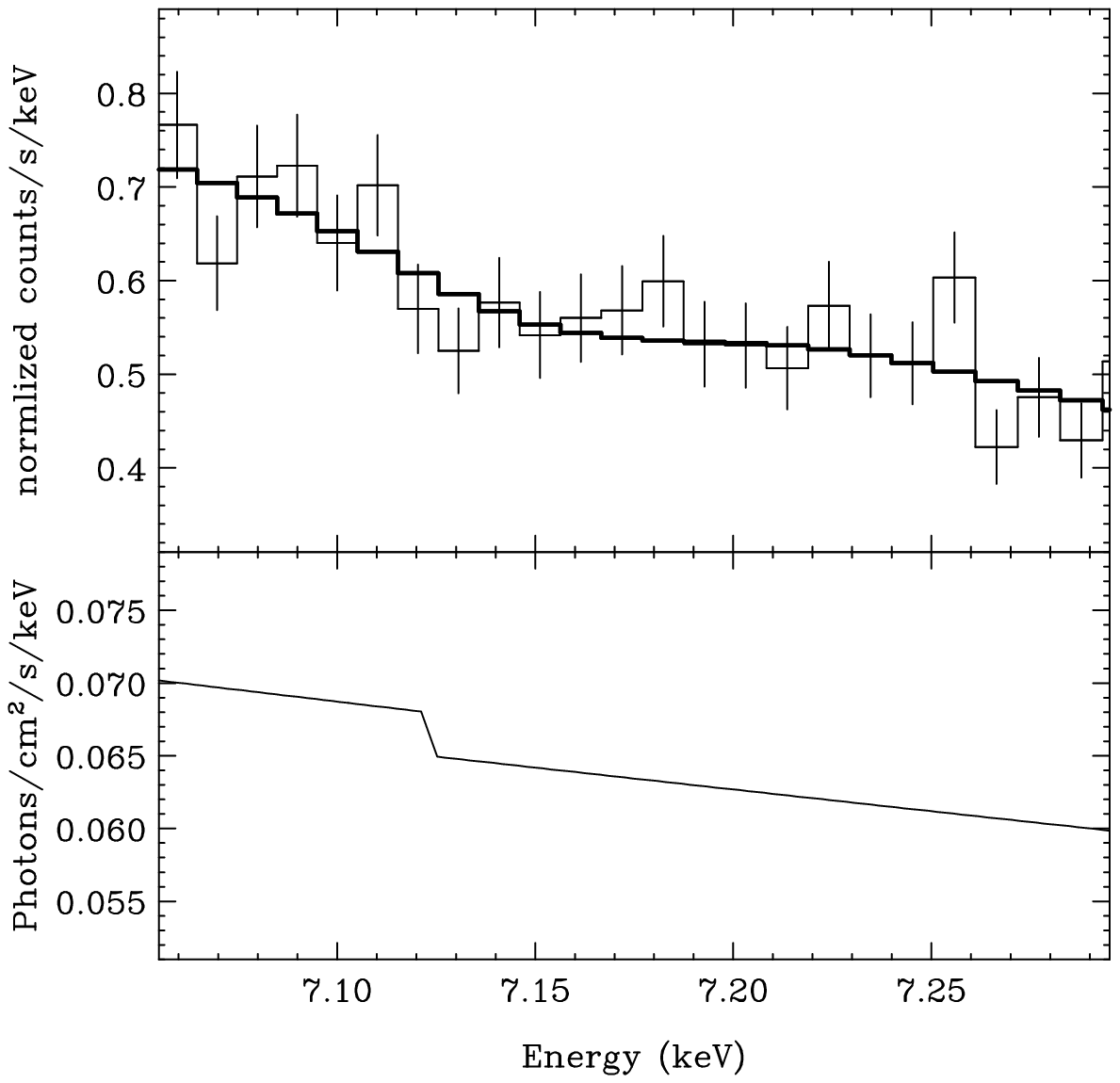}
\plotone{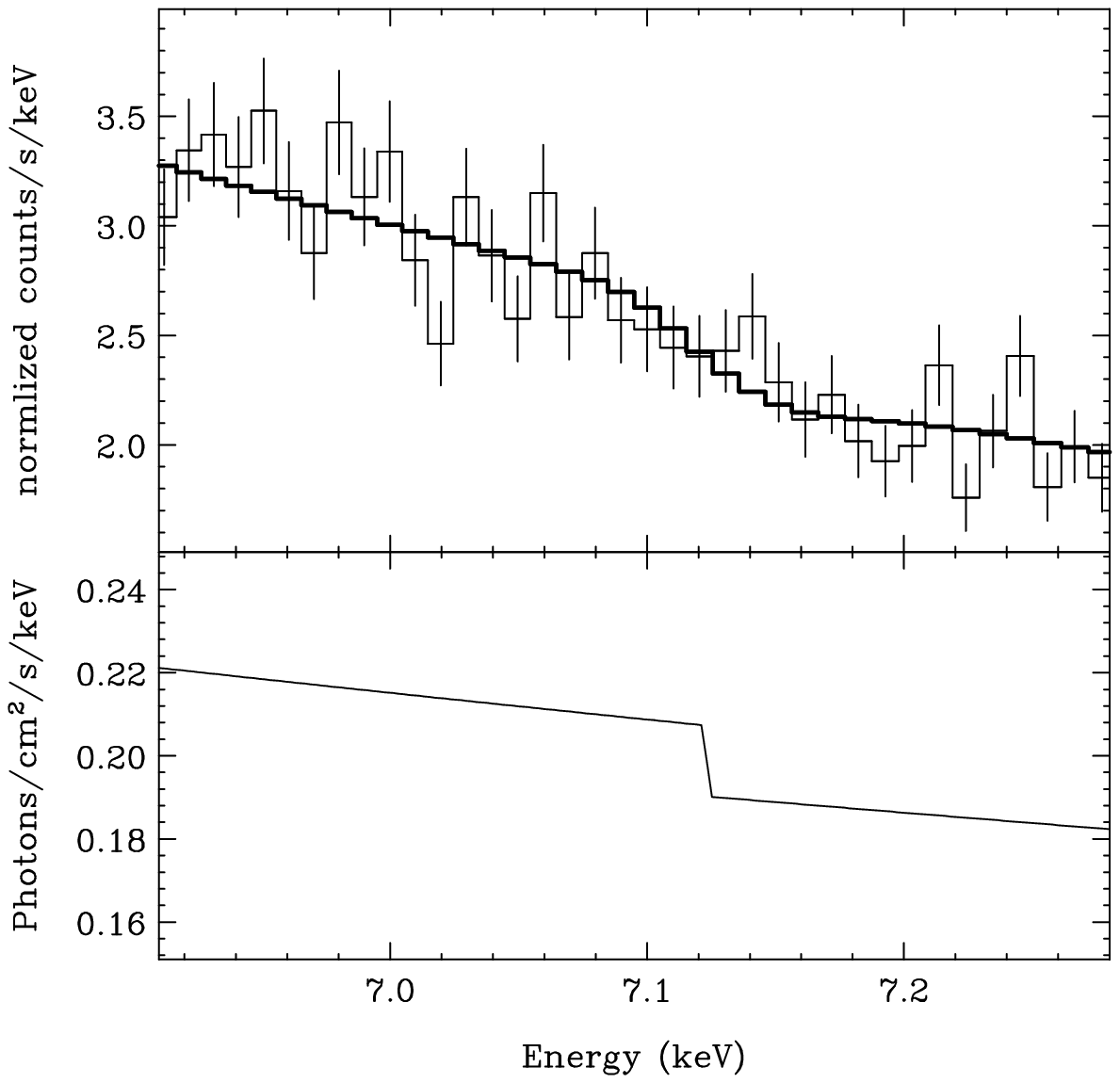}
\plotone{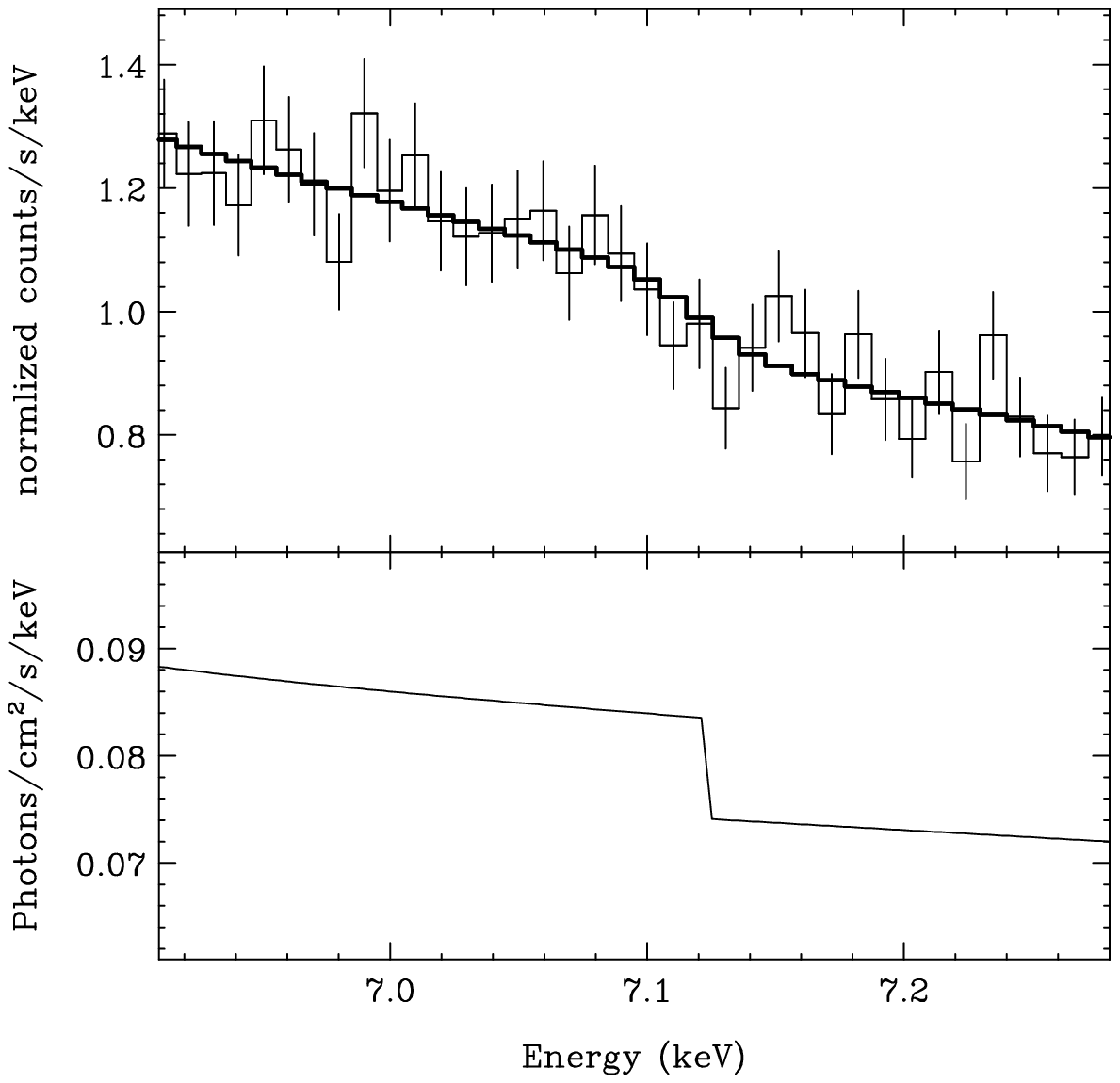}
\caption{
HEG spectra of (a) \gxp, (b) \gxm, and (c) \gxz, around the Fe 
K-band. The best-fit folded models ({\it TBvarabs} model) are overplotted onto
the data in the upper plots, while the unfolded models are in
the lower plots.
\label{fig5}}
\end{figure}

\fi

\clearpage
\begin{deluxetable}{ccccc}
\scriptsize
\tablenum{1}
\tablecaption{Multiwavelength Properties of the Targets\label{tbl-1}}
 \tablehead{\colhead{Target} &\colhead{Radio\tablenotemark{a}} &\colhead{$K$ Band\tablenotemark{b}} &\colhead{$N_{\rm H}$\tablenotemark{c}} &\colhead{Distance\tablenotemark{b}}\\
\colhead{} &\colhead{(mJy)} &\colhead{(mag)} &\colhead{($10^{22}$ cm$^{-2}$)} &\colhead{(kpc)}}
\startdata
GX~13+1  & 1.8$\pm$0.3 & $\sim$12	& 2.1 & 7$\pm1$ \\
GX~5--1  & 1.3$\pm$0.3 & 13.7		& 2.8 & 9.2$\pm$2.7 \\
GX~340+0 & 0.6$\pm0.3$ & 17.3		& 4.8 & 11.0$\pm$3.3 \\
\enddata
\tablenotetext{a}{The mean cm radio flux density taken from Table~4 of \citet{hom04}.}
\tablenotetext{b}{Taken from Table~4 of \citet{jon00}.}
\tablenotetext{c}{The hydrogen column density derived from the X-ray
spectral fit by assuming solar abundance ratios between H, He, C,
N, O, and Ne (this work; see Table~3).}
\end{deluxetable}

\begin{deluxetable}{ccccc}
\scriptsize
\tablenum{2}
\tablecaption{Observation Log\label{tbl-2}}
 \tablehead{\colhead{Target} &\colhead{Obs. ID} &\colhead{Start Time (UT)} &\colhead{End Time} &\colhead{Exposure}}
\startdata
GX~13+1  & 2708 &2002/10/08 11:27  & 19:54 & 29.4 ksec\\
GX~5--1  &  716 &2000/07/18 06:49 & 09:45 &  8.9 ksec\\
GX~340+0 & 1921 &2001/08/09 05:53 & 12:34 & 23.3 ksec\\
\enddata
\end{deluxetable}

\clearpage
\tabcolsep=2pt
\begin{deluxetable}{cccc}
\tablenum{3}
\tablecaption{Column Density of Abundant Elements and Continuum Parameters\label{tbl-3}}
\tablehead{\colhead{} &\colhead{GX~13+1} &\colhead{GX~5--1} &\colhead{GX~340+0} }
\startdata
\cutinhead{Column Densities ($10^{18}$ cm$^{-2}$) }
H  &21000$^{+34000}_{-13000}$&28000$^{+33000}_{-18000}$ &48000$^{+50000}_{-29000}$ \\
O  &18$^{+14}_{-7}$ &24$\pm13$ &40$\pm16$\\
Mg &1.81$\pm0.26$ &2.37$\pm0.33$ &3.5$^{+2.0}_{-2.4}$\\
Si &1.57$\pm0.11$ &1.83$\pm0.14$ &2.87$\pm0.28$\\
S  &1.02$\pm0.25$ &0.89$\pm0.20$ &1.25$\pm0.16$\\
Fe &1.3($<3.5$) &2.6$\pm1.4$ &3.5$\pm1.4$ \\
\cutinhead{Continuum Parameters\tablenotemark{a}}
$kT_{\rm db}$ &1.41$\pm$0.04  &1.04$\pm$0.05  &0.85$\pm$0.07  \\
 $F_{\rm db}$ &11  & 27 &  5.2\\ 
$kT_{\rm bb}$ &2.6 (fixed) &2.00$\pm$0.07  &1.91$\pm$0.05  \\
 $F_{\rm bb}$ &2.2  & 16 &  7.8\\
\enddata
\tablenotetext{a}{The parameters are obtained from the HEG first-order
spectrum at the best-fit column densities for the above elements
assuming solar abundance ratios within each group of H-He-C-N-O-Ne,
Na-Mg-Al, S-Cl-Ar-Ca, and Cr-Fe-Co-Ni. Dust scattering is
taken into account based on the \citet{dra03b} cross section.
Local spectral features 
(emission and absorption lines) are included in the fit (see text). 
The values $kT_{\rm 
db}$ and $kT_{\rm bb}$ are the innermost temperature of the MCD
model and the temperature of the blackbody component in keV, with
the extinction (absorption+scattering)
corrected 1--10 keV fluxes $F_{\rm db}$ and $F_{\rm bb}$ 
in units of $10^{-9}$ ergs cm$^{-2}$ s$^{-1}$, respectively.}
\tablecomments{The errors are 90\% confidence statistical errors except for the H and O column densities, for which systematic errors caused by the assumption of abundance ratios are considered (see text).}
\end{deluxetable}

\clearpage
\tabcolsep=2pt
\begin{deluxetable}{cccc}
\tablenum{4}
\tablecaption{K-Edge Energy of Free Atoms in the Literature\label{tbl-4}}
\tablehead{
\colhead{Mg} &\colhead{Si} &\colhead{S} &\colhead{Reference}\\
\colhead{(eV)} &\colhead{(eV)} &\colhead{(eV)} &\colhead{}}
\startdata
1311 & 1846 & 2477 	& 1 (E)\\
1310.6 &1848.6 & 2479.9 & 2 (T)\\
1311 &1846 & 2477 	& 3 (T)\\
1312.3 &1850.3 &2481.7	& 4 (T)\\
1313.8 &1853.0 &2485.5	& 5 (T)\\
1322.3(+) &1862.2(+) &2796.2(+) & 6 (T)\\
\enddata
\tablecomments{The value with (+) is that for a singly ionized ions (\ion{Mg}{2}, \ion{Si}{2}, and \ion{S}{2}).}
\tablecomments{References: (1) \citet{sev79}. (2) \citet{hua76}. (3) \citet{ver95}. (4) \citet{ind98}. (5) \citet{jun91}. (6) \citet{gou91}. E: Experimental, T: Theoretical.}
\end{deluxetable}

\clearpage
\tabcolsep=2pt
\begin{deluxetable}{cccccccc}
\tablenum{5}
\tablecaption{Summary of Abundance Ratios \label{tbl-5}}
\tablehead{\colhead{} &\colhead{GX~13+1} &\colhead{GX~5--1} &\colhead{GX~340+0} &\colhead{}&\multicolumn{3}{c}{Average\tablenotemark{a}}\\
\colhead{} &\colhead{(solar)} &\colhead{(solar)} &\colhead{(solar)} 
 &\colhead{}&\colhead{(solar)} &\colhead{(ISM)} &\colhead{(Number)}}
\startdata
${\rm H/Si}$ 
&0.48$^{+0.77}_{-0.29}$ &0.55$^{+0.64}_{-0.35}$ &0.59$^{+0.62}_{-0.36}$ 
&&0.55$\pm$0.39 &0.29$\pm$0.21 &$15000\pm11000$\\
${\rm O/Si}$ 
&0.48$^{+0.37}_{-0.17}$ &0.55$^{+0.29}_{-0.26}$ &0.59$^{+0.24}_{-0.22}$
&&0.55$\pm$0.17 &0.50$\pm$0.15 &13.3$\pm$3.9\\
${\rm Mg/Si}$ 
&1.08$\pm$0.17 &1.21$\pm$0.19 &1.15$^{+0.64}_{-0.78}$
&&1.14$\pm$0.13 &0.90$\pm$0.10 &1.22$\pm$0.14 \\
${\rm S/Si}$ 
&1.42$\pm0.36$ &1.06$\pm$0.25 &0.95$\pm0.15$ 
&&1.03$\pm$0.12 &0.71$\pm$0.09 &0.47$\pm$0.06 \\
${\rm Fe/Si}$ 
&0.64($<1.70$) &1.06$\pm$0.59 &0.94$\pm0.37$
&&0.97$\pm$0.31 &0.88$\pm$0.29 &1.28$\pm$0.41 \\
${\rm Mg/O}$ 
&2.2$^{+1.3}_{-1.1}$ &2.2$^{+2.0}_{-0.9}$ &2.0$\pm1.5$
&&2.2$\pm$1.1 &1.94$\pm$0.90 &0.100$\pm$0.046 \\
\enddata
\tablenotetext{a}{Given in 3 different units: relative to the solar abundance 
\citep{and89}, to the ``ISM'' abundances adopted by \citet{wil00}, and in number. The result of \gxp, \gxz, and \gxz\ is excluded to calculate the average of Fe/Si, Mg/Si, and Mg/O, respectively.}
\tablecomments{The errors are 90\% confidence level.}
\end{deluxetable}

\end{document}